\documentclass[12pt]{article}  
\usepackage{graphicx}
\usepackage{a41}
\usepackage{color}
\usepackage{cite}
\usepackage[rflt]{floatflt}

\usepackage{array}

\usepackage[english]{babel}
\usepackage[latin1]{inputenc}
\usepackage[T1]{fontenc}
\usepackage{ae}

\usepackage{url}

\usepackage{amsmath, amsthm, amssymb}
\newtheorem{thm}{Theorem}[section]

\newtheorem{definition}[thm]{Definition}

\newcommand{\ppgq}{{p}_{gq}}
\newcommand{\pgq}{\bar{p}_{gq}}
\newcommand{\MS}{\overline{{\sf MS}}}

\newcommand{\N}{\nonumber}

\newcommand{\ep}{\varepsilon}

\usepackage{rotating}

\usepackage{algorithm}
\usepackage{algorithmicx}
\usepackage{algpseudocode}
\usepackage{graphicx}

\newcounter{mmacnt}
\def\restartmma{\setcounter{mmacnt}{0}}
\restartmma \catcode`|=\active
\def|#1|{\mathrm{#1}}
\catcode`|=12
\newenvironment{mma}{
 \par\smallskip
 \catcode`|=\active
 \parskip=0pt\parindent=0pt 
 \small
 \def\In##1\\{%
   \def\linebreak{\hfill\break\null\qquad}%
   \refstepcounter{mmacnt}
   \hangindent=2.5em\hangafter=0
   \leavevmode
   \llap{\tiny\sffamily In[\arabic{mmacnt}]:=\kern.5em}%
   \mathversion{bold}\footnotesize$\displaystyle##1$\normalsize
   \mathversion{normal}\par
 }%
 \def\Print##1\\{%
   \def\linebreak{\hfill\break}%
   \hangindent=2.5em\hangafter=0
   \leavevmode ##1\par}%
 \def\Out##1\\{%
   \def\linebreak{$\hfill\break\null\hfill$}%
   \kern\abovedisplayskip\par
   \hangindent=2.5em\hangafter=0
   \leavevmode
   \llap{\tiny\sffamily Out[\arabic{mmacnt}]=\kern.5em}
   \footnotesize$\displaystyle##1$\normalsize\hfill\null\par
   \kern\belowdisplayskip
 }%
 \def\Warning##1##2\\{%
   \def\linebreak{\hfill\break}%
   \hangindent=2.5em\hangafter=0
   \leavevmode
   {\scriptsize##1 : ##2}\par}%
}{%
 \par\smallskip
}

\usepackage{color}

\newenvironment{fshaded}{%
\MakeFramed {\FrameRestore}
}%
{\endMakeFramed}

\usepackage{tikz}
\usetikzlibrary{matrix}

\allowdisplaybreaks[4]

\begin{document}
\setlength{\baselineskip}{0.515cm}
\sloppy
\thispagestyle{empty}
\begin{flushleft}
DESY 13--185
\\
DO-TH 13/27\\
MITP/13-063\\
SFB/CPP-13-74\\
LPN 13-072\\ 
Higgstools 14-001\\ 
January 2014\\
\end{flushleft}

\mbox{}
\vspace*{\fill}
\begin{center}

{\LARGE\bf The Transition Matrix Element \boldmath{$A_{gq}(N)$} of the} 

\vspace*{1mm}
{\LARGE\bf Variable Flavor Number Scheme at \boldmath{$O(\alpha_s^3)$}}

\vspace{4cm}
\large
J.~Ablinger$^a$,  
J.~Bl\"umlein$^b$, 
A.~De Freitas$^b$, 
A.~Hasselhuhn$^a$, 
A.~von~Manteuffel$^c$,
M.~Round$^{a,b}$,
C.~Schneider$^a$ and
F.~Wi\ss{}brock$^{a,b}$\footnote{Present address: IHES, 35 Route de Chartres, 91440 Bures-sur-Yvette, 
France.}

\vspace{1.5cm}
\normalsize
{\it $^a$~Research Institute for Symbolic Computation (RISC),\\
                          Johannes Kepler University, Altenbergerstra\ss{}e 69,
                          A--4040, Linz, Austria}\\

\vspace*{3mm}
{\it  $^b$ Deutsches Elektronen--Synchrotron, DESY,}\\
{\it  Platanenallee 6, D-15738 Zeuthen, Germany}

\vspace*{3mm}
{\it  $^c$ PRISMA Cluster of Excellence and Institute of Physics, J. Gutenberg University,}\\
{\it D-55099 Mainz, Germany.}
\\

\end{center}
\normalsize
\vspace{\fill}
\begin{abstract}
\noindent 
We calculate the massive operator matrix element $A_{gq}^{(3)}(N)$ to 3-loop order
in Quantum Chromodynamics at general values of the Mellin variable $N$. This is the first
complete transition function needed in the variable flavor number scheme obtained at  
$O(\alpha_s^3)$. A first independent recalculation is performed for the contributions 
$\propto N_F$ of the 3-loop anomalous dimension $\gamma_{gq}^{(2)}(N)$.
\end{abstract}

\vspace*{\fill}
\noindent
\numberwithin{equation}{section}
\newpage

\section{Introduction}
\label{sec:1}

\vspace*{1mm}
\noindent
The variable flavor number scheme (VFNS) allows, in a process independent manner, the transition of the twist-2 
parton distributions for $N_f$ light flavors to $N_f + 1$ light flavors at a scale $\mu^2$, i.e. making one single 
heavy flavor $Q$ light at the time. This has been worked out to 2-loop order in \cite{Buza:1996wv} and to 3-loop order 
in \cite{Bierenbaum:2009mv}. The new $(N_f+1)$-flavor massless parton densities for the different flavor 
combinations are given by~: 
\begin{eqnarray}
  {f_k(N_f+1, \mu^2) + f_{\overline{k}}(N_f+1, \mu^2)} 
&=& {A_{qq,Q}^{\rm NS} \Big(N_f, \frac{\mu^2}{m^2}\Big)}\otimes
  {\left[f_k(N_f, \mu^2) + f_{\overline{k}}(N_f, \mu^2)\right]} \nonumber\\
& & + {\tilde A_{qq,Q}^{\rm PS}\Big(N_f, \frac{\mu^2}{m^2}\Big)}\otimes
  {\Sigma(N_f, \mu^2)} 
\nonumber\\ &&
 + {\tilde A_{qg,Q}^{\rm S}\Big(N_f, \frac{\mu^2}{m^2}\Big)}\otimes
  {G(N_f, \mu^2)},
\nonumber\\
  {f_{Q+\bar Q}(N_f+1, \mu^2)} 
   &=& \tilde A_{Qq}^{\rm PS}\Big(N_f, \frac{\mu^2}{m^2}\Big)\otimes
       {\Sigma(N_f, \mu^2)}
   + \tilde A_{Qg}^{\rm S}\Big(N_f, \frac{\mu^2}{m^2}\Big) \otimes
    {G(N_f, \mu^2)},
    \N  \\
 {G(N_f+1, \mu^2)} &=&  {A_{gq,Q}^{\rm S}\Bigl(N_f,\frac{\mu^2}{m^2}\Bigr)}
\otimes {\Sigma(N_f,\mu^2)}
+ A_{gg,Q}^S\Bigl(N_f,\frac{\mu^2}{m^2}\Bigr)\otimes {G(N_f,\mu^2)},  
\N\\
{\Sigma(N_f+1,\mu^2)} &=&
{\sum_{k=1}^{N_f+1} \left[f_k(N_f+1,\mu^2) + 
f_{\overline{k}}(N_f+1,\mu^2)\right]} \nonumber\\
&=& \left[{A_{qq,Q}^{\rm NS}\Big(N_f, \frac{\mu^2}{m^2}\Big)} + 
N_f 
{\tilde{A}_{qq,Q}^{\rm PS} \Big(N_f, \frac{\mu^2}{m^2}\Big)} +
\tilde{A}_{Qq}^{\rm PS} \Big(N_f, \frac{\mu^2}{m^2}\Big) \right]
\nonumber\\ &&
\otimes {\Sigma(N_f,\mu^2)} \nonumber\\ &&
+\left[N_f 
{\tilde{A}^{\rm S}_{qg,Q}\Big(N_f, \frac{\mu^2}{m^2}\Big)} 
+ \tilde{A}^{\rm S}_{Qg}\Big(N_f, \frac{\mu^2}{m^2}\Big) \right]
\otimes {G(N_f,\mu^2)} .
\end{eqnarray}
Here, $f_k$ and $f_{\bar{k}}$ denote the quark and antiquark densities, $\Sigma(N_f, \mu^2) 
= \sum_{k=1}^{N_f} [f_k + f_{\bar{k}}]$ is the singlet-quark density and $G(N_f, \mu^2)$ 
the gluon density. The mass $m$ of the decoupling heavy quark $Q$, enters the massive 
operator matrix elements (OMEs) $A_{ij}(N_F, m^2/\mu^2)$, which are process independent 
quantities, in terms of logarithms, and $\mu$ denotes the decoupling scale. 
In total seven OMEs contribute. Here we use the shorthand 
notations $\tilde{f} = f/N_F,~~\hat{f} = f(N_F+1) - f(N_F)$. Representations of parton 
distribution functions in the VFNS are important at very large virtualities, as e.g. for 
scattering processes at the Tevatron or the Large Hadron Collider (LHC), in particular in the context
of precision measurements of different observables in Quantum Chromodynamics (QCD) and the determination
of the strong coupling constant $\alpha_s(M_Z)$ \cite{Bethke:2011tr}.
 At 1--loop order 
only 
the OME $\tilde A_{Qg}^{\rm S}$ contributes. At 2--loop order $ \tilde A_{Qq}^{\rm PS}, 
A_{qq,Q}^{\rm NS},  A_{gg,Q}^S, A_{gq,Q}^{\rm S}$ emerge, and from 3--loop order onward also $\tilde A_{qq,Q}^{\rm PS}$ 
and $\tilde A_{qg,Q}^{\rm S}$ contribute. The 2-loop corrections to all the matrix elements have been calculated in 
Refs.~\cite{Buza:1995ie,Buza:1996wv,Bierenbaum:2007qe,Bierenbaum:2009zt} in complete form. At 3--loop order, 
Mellin moments ranging for $N = 2 ... 10(14)$, depending on the process, were calculated in \cite{Bierenbaum:2009mv} and 
for transversity $A_{qq,Q}^{\rm NS,TR}$ in \cite{Blumlein:2009rg}.

The OMEs $\tilde A_{qq,Q}^{\rm PS}(N)$ and $\tilde A_{qg,Q}^{\rm S}(N)$ have been computed in 
\cite{Ablinger:2010ty}. There also the 3--loop $O(N_f T_F^2 C_{A,F})$ corrections to the OMEs
to $\tilde A_{Qg}^{\rm S}, \tilde A_{Qq}^{\rm PS}, A_{qq,Q}^{\rm NS}$  and $A_{qq,Q}^{\rm NS,TR}$
were calculated, with $T_F = 1/2, C_A = N_c, C_F = (N_c^2-1)/(2 N_c)$ for the gauge group $SU(N_c)$. The corresponding 
contributions to $A_{gg,Q}^S$ and $A_{gq,Q}^{\rm S}$ were calculated in 
\cite{Blumlein:2012vq}.

In the present paper we compute the complete matrix elements $A_{gq,Q}^{\rm S (3)}(N)$ for 
general values of
$N$. As a by-product of the calculation we also obtain the two-loop anomalous dimension $\gamma_{gq}^{(1)}(N)$ 
and the term $\propto T_F$ of the three-loop anomalous dimension, $\bar{\gamma}_{gq}^{(2)}(N)$. The paper is 
organized 
as follows. We first describe technical details of the calculation and then present the constant part of 
the massive 3-loop operator matrix element $A_{gq,Q}^{(3) \rm S}$ in Mellin-$N$ space as well as the 
$\bar{\gamma}_{gq}^{(2)}(N)$ which is obtained in an independent calculation. In the Appendix we present the matrix 
element $A_{gq,Q}^{\rm S}$ in Mellin- and momentum fraction 
space.

\section{The Formalism}
\label{sec:2}

\vspace*{1mm}
\noindent
The massive operator matrix element $A_{gq,Q}^{(3), \MS}$ in the $\MS$-scheme is given by
\cite{Bierenbaum:2009mv}~:
\begin{eqnarray}
    A_{gq,Q}^{(3), \MS}&=&
                     -\frac{\gamma_{gq}^{(0)}}{24}
                      \Biggl\{
                          \gamma_{gq}^{(0)}\hat{\gamma}_{qg}^{(0)}
                         +\Bigl(
                              \gamma_{qq}^{(0)}
                             -\gamma_{gg}^{(0)}
                             +10\beta_0
                             +24\beta_{0,Q}
                                     \Bigr)\beta_{0,Q}
                      \Biggr\}
                           \ln^3 \Bigl(\frac{m^2}{\mu^2}\Bigr)
                     +\frac{1}{8}\Biggl\{
                         6\textcolor{blue}{\gamma_{gq}^{(1)}}\beta_{0,Q}
\N\\ &&
                        +\hat{\gamma}_{gq}^{(1)}\Bigl(
                                      \gamma_{gg}^{(0)}
                                     -\gamma_{qq}^{(0)}
                                     -4\beta_0
                                     -6\beta_{0,Q}
                                                 \Bigr)
                        +\gamma_{gq}^{(0)}\Bigl(
                                       \hat{\gamma}_{qq}^{(1), {\sf NS}}
                                      +\hat{\gamma}_{qq}^{(1), {\sf PS}}
                                      -\hat{\gamma}_{gg}^{(1)}
                                      +2\beta_{1,Q}
                                                 \Bigr)
                      \Biggr\}
                           \ln^2 \Bigl(\frac{m^2}{\mu^2}\Bigr)
\N\\ &&
                     +\frac{1}{8}\Biggl\{
                              4 \textcolor{blue}{\hat{\gamma}_{gq}^{(2)}}
                            + 4a_{gq,Q}^{(2)}         \Bigl(
                                    \gamma_{gg}^{(0)}
                                   -\gamma_{qq}^{(0)}
                                   -4\beta_0
                                   -6\beta_{0,Q}
                                                       \Bigr)
                            + 4\gamma_{gq}^{(0)}       \Bigl(
                                      a_{qq,Q}^{(2),{\sf NS}}
                                     +a_{Qq}^{(2),{\sf PS}}
                                     -a_{gg,Q}^{(2)}
\N\\ &&
                                     +\beta_{1,Q}^{(1)}
                                                       \Bigr)
                            + \gamma_{gq}^{(0)}\zeta_2 \Bigl(
                               \gamma_{gq}^{(0)}\hat{\gamma}_{qg}^{(0)}
                               +\Bigl[
                                        \gamma_{qq}^{(0)}
                                       -\gamma_{gg}^{(0)}
                                       +12\beta_{0,Q}
                                       +10\beta_0
                                            \Bigr]\beta_{0,Q}
                                                       \Bigr)
                      \Biggr\}
                           \ln \Bigl(\frac{m^2}{\mu^2}\Bigr)
\N\\ &&
                  + \overline{a}_{gq,Q}^{(2)} \Bigl(
                                       \gamma_{qq}^{(0)}
                                      -\gamma_{gg}^{(0)}
                                      +4\beta_0
                                      +6\beta_{0,Q}
                                             \Bigr)
                  + \gamma_{gq}^{(0)} \Bigl(
                                       \overline{a}_{gg,Q}^{(2)}
                                      -\overline{a}_{Qq}^{(2),{\sf PS}}
                                      -\overline{a}_{qq,Q}^{(2),{\sf NS}}
                                             \Bigr)
                -\gamma_{gq}^{(0)}\beta_{1,Q}^{(2)}
\N\\ && 
                -\frac{\gamma_{gq}^{(0)}\zeta_3}{24} \Bigl(
                           \gamma_{gq}^{(0)}\hat{\gamma}_{qg}^{(0)}
                          +\Bigl[
                                   \gamma_{qq}^{(0)}
                                  -\gamma_{gg}^{(0)}
                                  +10\beta_0
                                      \Bigr]\beta_{0,Q}
                                             \Bigr)
                -\frac{3\gamma_{gq}^{(1)}\beta_{0,Q}\zeta_2}{8}
                +2 \delta m_1^{(-1)} a_{gq,Q}^{(2)}
\N\\ &&
                +\delta m_1^{(0)} \hat{\gamma}_{gq}^{(1)}
                +4 \delta m_1^{(1)} \beta_{0,Q} \gamma_{gq}^{(0)}
                +a_{gq,Q}^{(3)}~, \label{Agq3QMSren}
\label{AgqQ3Ren1}
\end{eqnarray}
where $\zeta_k = \sum_{l=1}^\infty (1/l^k),~~~k \in \mathbb{N}, k \geq 2$ are the values of the Riemann 
$\zeta$-function. 
The different logarithmic terms $\propto \ln^n(m^2/\mu^2)$, from highest to lowest power $n =3 ...0$, depend 
on anomalous dimensions 
$\gamma_{ij}^{(k)}(N), (k = 0...2)$, the expansion coefficients of the QCD $\beta$-function, the  heavy quark mass, 
and 
of the constant parts of the unrenormalized massive OMEs $a_{ij}^{(k)}$ 
\cite{Buza:1995ie,Bierenbaum:2007qe,Buza:1996wv,Bierenbaum:2009zt,Bierenbaum:2008yu} from 1- to 3-loop 
order. Calculating the OMEs order by order in $\alpha_s$ delivers all these quantities. Moreover, one obtains from 
the $\ln^2(m^2/\mu^2)$ term the complete 2-loop anomalous dimension $\gamma_{gq}^{(1)}(N)$ and from the contribution 
$\ln(m^2/\mu^2)$ the term $\propto T_F$ of the 3-loop anomalous dimension $\gamma_{gq}^{(2)}(N)$. 

The 86 Feynman diagrams contributing to  $A_{gq,Q}^{(3)}$ were generated with an extension of {\tt QGRAF}  
\cite{Nogueira:1991ex} allowing for local operator insertions \cite{Bierenbaum:2009mv}. The diagrams are 
calculated using {\tt TFORM} \cite{Tentyukov:2007mu}. With respect to the latter the corresponding diagrams  are 
mapped into generating functions in a subsidiary variable $x$, cf.~\cite{Ablinger:2012qm}. 
The resulting Feynman integrals contain not only the usual propagator-denominators, but also denominator factors
in which the loop momenta enter linearly. The latter stem from the local operator insertions which have been 
resummed in terms of a generating function representation. 
We then use  
integration-by-parts relations \cite{IBP}  as encoded in {\tt Reduze2} \cite{vonManteuffel:2012np}, which has 
been 
adapted to this extension. The master integrals are calculated using hypergeometric function techniques 
\cite{GHYP,Slater,Appell,Hamberg,Bierenbaum:2007qe,Ablinger:2012qm} and Mellin-Barnes \cite{MELB} representations 
applying the codes {\tt MB} \cite{Czakon:2005rk} and {\tt MBresolve} \cite{Smirnov:2009up}. One obtains multiple 
nested 
sums over hypergeometric expressions still containing the dimensional parameter $\ep = D - 4$. After expanding in 
$\varepsilon$, these sums are performed applying modern summation technologies
\cite{Karr:81,Schneider:01,Schneider:05a,Schneider:07d,Schneider:08c,
Schneider:10a,Schneider:10b,Schneider:10c,Schneider:13b} encoded in the packages {\tt Sigma} \cite{SIG1,SIG2}, {\tt 
HarmonicSums} \cite{HARMONICSUMS,Ablinger:2010kw}, 
{\tt EvaluateMultiSums}, {\tt SumProduction} \cite{EMSSP}, and {\tt $\rho$-Sum} \cite{RHOSUM}. All but one master 
integral could be calculated in this way. For the missing case we have applied the multivariate 
Almkvist-Zeilberger algorithm \cite{AZ} in Mellin-space, which allowed us to find a difference equation using the 
package
{\tt MultiIntegrate} \cite{Ablinger:2010kw}. This equation could then be solved applying the summation packages quoted 
before. Both the master integrals as well as the final results for individual diagrams were checked 
comparing to a finite number of moments calculated using {\tt MATAD} \cite{Steinhauser:2000ry}.

Finally it turns out that the OME $A_{gq,Q}^{(3)}$ can be expressed by harmonic sums $S_{\vec{a}}(N)$ and 
$\zeta$-values \cite{Blumlein:2009cf} only. The harmonic sums are defined by \cite{HSUM}
\begin{eqnarray}
S_{b,\vec{a}}(N) = \sum_{k=1}^N \frac{{\rm sign}(b)^k}{k^{|b|}} S_{\vec{a}}(k),~~~S_{\emptyset}(N) = 1; a_i,b \in 
\mathbb{Z} \setminus \{0\}, N \in \mathbb{N} \setminus \{0\}.
\end{eqnarray}

The renormalization and factorization of $A_{gq,Q}^{(3)}$ has been worked out in 
Ref.~\cite{Bierenbaum:2009mv}. It consists of mass, coupling constant and operator renormalization, 
and the factorization of the collinear singularities. Unlike the case of massless OMEs, the $Z$-factors 
for the renormalization of the ultraviolet singularities of the operators are not inverse of those
for the collinear singularities. For the renormalization  of the coupling constant, one first
refers to a MOM-scheme, using the background-field method \cite{BGF} and then translates
to the $\MS$-scheme afterwards.

\section{Anomalous Dimensions}
\label{sec:3}

\vspace*{1mm}
\noindent
The anomalous dimension may be obtained from the logarithmic contributions in Eq.~(\ref{AgqQ3Ren1}).
In the following we drop the argument $N$ of the harmonic sums and use the short-hand notation 
$S_{\vec{a}}(N) \equiv S_{\vec{a}}$. Here and in the following we simplify the result applying the 
algebraic relations of the harmonic sums \cite{Blumlein:2003gb}.
In the present calculation the complete 2-loop 
anomalous dimension $\gamma_{gq}^{(1)}(N)$ and the $T_F$-part of the 3-loop anomalous dimension
$\bar{\gamma}_{gq}^{(2)}(N)$ are calculated in an independent way and may be compared to the
result obtained previously in Ref.~\cite{Vogt:2004mw}.

Let us define the leading order splitting function $\bar{p}_{gq}(N)$ without color factor,
\begin{eqnarray}
\bar{p}_{gq}(N) = \frac{N^2 + N  +2}{(N-1)N(N+1)}~. 
\end{eqnarray}
The 2--loop anomalous dimension $\gamma_{gq}^{(1)}$ is then given by
\begin{eqnarray}
\gamma_{gq}^{(1)}(N) &=& \frac{1}{2}[1 + (-1)^N] \Biggl\{
\textcolor{blue}{C_A C_F} \Biggl[
8 \bar{p}_{gq} [S_2 - S_1^2 + 2 S_{-2}]
+\frac{8 \big(17 N^4+41 N^2-22 N-12\big)}{3 (N-1)^2 N^2 (N+1)} S_1
\nonumber\\ &&
-(-1)^N \frac{16 \big(3 N^3+5 N^2+6 N+2\big)}{(N-1) N^2 (N+1)^3}
-\frac{8 Q_1}
      {9 (N-1)^2 N^3 (N+1)^2 (N+2)^2}\Biggr]
\nonumber\\ &&
+ \textcolor{blue}{C_F^2} \Biggl[
8 \bar{p}_{gq}[S_1^2 + S_2]
-\frac{8 \big(5 N^3+8 N^2+17 N+10\big)}{(N-1) N (N+1)^2} S_1
+\frac{4 Q_2}
{(N-1) N^3 (N+1)^3}
\Biggr]
\nonumber\\ &&
+ \textcolor{blue}{C_F T_F N_F} \Biggl[
\frac{32 \big(8 N^3+13 N^2+27 N+16\big)}{9 (N-1) N (N+1)^2}
-\frac{32}{3} \bar{p}_{gq} S_1
\Biggr]\Biggr\},
\end{eqnarray}
where $Q_1$ and $Q_2$ are the polynomials 
\begin{eqnarray}
Q_1 &=& 109 N^8+512 N^7+834 N^6+592 N^5-275 N^4-900 N^3-152 N^2+432 N+144,
\nonumber\\ &&
\\
Q_2 &=& 12 N^6+30 N^5+43 N^4+28 N^3-N^2-12 N-4~.
\end{eqnarray}
The contribution to the 3-loop anomalous dimension is given by
\begin{eqnarray}
\bar{\gamma}_{gq}^{(2)}(N) &=& \frac{1}{2}\left[1 + (-1)^N\right]\Biggl\{
\textcolor{blue}{C_F^2 T_F N_F} \Biggl\{
 \frac{80}{9} \bar{p}_{gq} S_1^3
-\frac{16 \big(58 N^4+95 N^3+192 N^2+113 N-6\big)}
      {9 (N-1) N^2 (N+1)^2} S_1^2
\nonumber\\ &&
+\Biggl[
\frac{64 Q_3}
     {27 (N-1) N^3 (N+1)^3}
+\frac{208}{3} \bar{p}_{gq} S_2
\Biggr] S_1
-(-1)^N \frac{64 Q_4}{9 (N-1)^2 N^2 (N+1)^4 (N+2)^3}
\nonumber\\ &&
+\frac{2 Q_5}
{27 (N-1)^2 N^5 (N+1)^5 (N+2)^3}
-\frac{16 \big(88 N^4+155 N^3+294 N^2+185 N+18\big)}{9 (N-1) N^2 (N+1)^2} S_2
\nonumber\\ &&
+\frac{256}{9} \bar{p}_{gq} S_3
-\frac{256}{(N-1) N (N+1) (N+2)} \bar{p}_{gq} S_{-2}
-\frac{64}{3} \bar{p}_{gq} \left[S_{2,1} +6 \zeta_3 \right]
\Biggr\} 
\nonumber\\ &&
+ \textcolor{blue}{C_F T_F^2 N_F^2} \Biggl\{
-\frac{32}{3} \bar{p}_{gq} [S_1^2 + S_2]
+\frac{64 \big(8 N^3+13 N^2+27 N+16\big)}{9 (N-1) N (N+1)^2} S_1
\nonumber\\ &&
-\frac{64 \big(4 N^4+4 N^3+23 N^2+25 N+8\big)}{9 (N-1) N (N+1)^3}
\Biggr\}
+ \textcolor{blue}{C_F C_A T_F N_F} \Biggl\{
-\frac{80}{9} \bar{p}_{gq} S_1^3
\nonumber\\ &&
+\frac{32 Q_6}
      {9 (N-1)^2 N^2 (N+1)^2 (N+2)} S_1^2
+\Biggl[
-\frac{16 Q_7}
{27 (N-1)^2 N^3 (N+1)^3 (N+2)^2}
\nonumber\\ &&
+\frac{80}{3} \bar{p}_{gq}  S_2
\Biggr] S_1
+\frac{8 Q_8}
{27 (N-1)^2 N^4 (N+1)^4 (N+2)^3}
+ 
\Biggl[
-\frac{32 Q_9}
{3 (N-1)^2 N^2 (N+1)^2 (N+2)}
\nonumber\\ &&
+\frac{256}{3} \bar{p}_{gq} S_1 \Biggr] S_{-2}
+(-1)^N \Biggl[\frac{32 Q_{10}}
{9 (N-1)^2 N^3 (N+1)^4 (N+2)^3}
\nonumber\\ &&
-\frac{128 \big(3 N^3+5 N^2+6 N+2\big)}{3 (N-1) N^2 (N+1)^3} S_1 \Biggr] 
-\frac{32 \big(N^6-N^4-7 N^3-20 N^2+5 N+10\big)}{3 (N-1)^2 N^2 (N+1)^2 (N+2)} S_2
\nonumber\\ &&
+ \frac{128}{3} \bar{p}_{gq}  \left[\frac{4}{3} S_3 +S_{-3} - S_{-2,1} + 3 \zeta_3\right]
\Biggl\} \Biggr\},
\end{eqnarray}
and the polynomials $Q_i$ are
\begin{eqnarray}
Q_3 &=& 88 N^6+171 N^5+374 N^4+320 N^3+23 N^2-78 N-18,
\\
Q_4 &=& 4 N^8+106 N^7+653 N^6+1411 N^5+453 N^4-2429 N^3-3394 N^2 -1644 N-344,
\\
Q_5 &=& 129 N^{14}+3522 N^{13}+27035 N^{12}+113832 N^{11}+320915 N^{10}+585194 N^9
        +526437 N^8
\nonumber\\ &&
-136068 N^7-910532 N^6-1191312 N^5-976400 N^4-562880 N^3-276672 N^2
\nonumber\\ &&
       -122112 N-27648,
\\
Q_6 &=& 32 N^6+96 N^5+132 N^4+167 N^3-26 N^2-167 N-54,
\\
Q_7 &=& 736 N^9+4233 N^8+10139 N^7+15625 N^6+15401 N^5+2050 N^4-10784 N^3
\nonumber\\ &&
-7400 N^2 -336 N+576,
\\
Q_8 &=& 1485 N^{12}+12021 N^{11}+40816 N^{10}+77198 N^9+77813 N^8+6809 N^7
-58634 N^6
\nonumber\\ &&
+5012 N^5+124920 N^4+145680 N^3+77984 N^2+18240 N+1152,
\\
Q_9 &=& 13 N^6+37 N^5+51 N^4+39 N^3-76 N^2-72 N-40,
\\
Q_{10} &=&
73 N^9+655 N^8+2495 N^7+4747 N^6+3420 N^5-2846 N^4-7048 N^3
\nonumber\\ &&
-4872 N^2-1616 N-192~.
\end{eqnarray}
Both quantities agree with the results in the literature for individual moments \cite{ANDIM3,Bierenbaum:2009mv} and 
the result for general values of $N$ \cite{ANDIM2}. In the 3-loop case the results given in \cite{Vogt:2004mw} are 
confirmed for the first time for general values of $N$. 

\section{The Operator Matrix Element}
\label{sec:4}

\vspace*{1mm}
\noindent
The renormalized operator matrix element~(\ref{AgqQ3Ren1}) is represented by known lower order terms
and the newly evaluated constant part $a_{gq}^{(3)}(N)$ of the unrenormalized OME~:  
\begin{eqnarray}
a_{gq}^{(3)}(N) &=& \frac{1}{2} [1 + (-1)^N] \Biggl\{
\textcolor{blue}{C_F^2 T_F} 
\Biggl\{
\bar{p}_{gq} \left(\frac{64}{3} B_4 - 96  \zeta_4\right)
-2 \Biggl[-\frac{29}{27} \bar{p}_{gq} S_1^4
\nonumber\\ &&
+ \frac{2 \big(275 N^4+472 N^3+951 N^2+598 N+96\big)}{81 (N-1) N^2 (N+1)^2} S_1^3
+\Biggl[
-\frac{2 P_1}{81 (N-1) N^3 (N+1)^3}
\nonumber\\  &&
+ \frac{14}{9} \bar{p}_{gq} S_2
\Biggr] S_1^2
+\Biggl[
-\frac{2 \big(209 N^3+376 N^2+669 N+418\big)}{27 (N-1) N (N+1)^2} S_2
-\frac{4 P_0}{243 (N-1) N^4 (N+1)^4}
\nonumber\\
&&
+\frac{104}{27} \bar{p}_{gq} S_3
- \frac{16}{9} \bar{p}_{gq} S_{2,1}
\Biggr] S_1
+\frac{1}{3} \bar{p}_{gq} S_2^2
+\frac{2 P_2}{243 (N-2) (N-1)^2 N^5 (N+1)^5 (N+2)^4}
\nonumber\\ &&
+\frac{2 P_3}{81 (N-2) (N-1)^2 N^4 (N+1)^4 (N+2)^2}  S_2
-\frac{64 \bar{p}_{gq} }{(N-1) N (N+1) (N+2)} S_{-1} S_2
\nonumber\\ &&
-\frac{4 P_4}{81 (N-1)^2 N^3 (N+1)^3 (N+2)} S_3
+\frac{110}{9}  \bar{p}_{gq} S_4
+\Biggl[
\frac{64 \bar{p}_{gq} }{(N-1) N (N+1) (N+2)} S_{-1} 
\nonumber\\ &&
+\frac{16 P_5}{3 (N-2) (N-1)^2 N^3 (N+1)^3 (N+2)^2}
\Biggr] S_{-2}
-\frac{64 \bar{p}_{gq} }{3 (N-1) N (N+1) (N+2)}\left[ S_{-3} \right. 
\nonumber\\ && \left.
- 3S_{2,-1} + 3 S_{-2,-1}\right]
+\frac{8 \big(35 N^3+64 N^2+111 N+70\big)}{27 (N-1) N (N+1)^2} S_{2,1}
-\frac{16}{9} \bar{p}_{gq} \left[3 S_{3,1} - S_{2,1,1}\right]
\Biggr]
\nonumber\\ &&
-2 
\Biggl[
\frac{2 \big(17 N^4+28 N^3+69 N^2+46 N+24\big)}{9 (N-1) N^2 (N+1)^2} S_1
+ \frac{P_6}{9 (N-1)^2 N^3 (N+1)^3 (N+2)^2}
\nonumber\\ &&
- \frac{1}{3} \bar{p}_{gq} \left( 10 S_1^2 - 14 S_2\right)
\Biggr] \zeta_2
+2 \Biggl[
\frac{2 P_7}{9 (N-1)^2 N^3 (N+1)^3 (N+2)}
+\frac{152}{9} \bar{p}_{gq} S_1
\Biggr]
\zeta_3
\Biggr\}
\nonumber\\ &&
+ \textcolor{blue}{C_F T_F^2} 
\Biggl\{
-2 \textcolor{blue}{N_F}
\Biggl[
 \frac{8}{27} \bar{p}_{gq} S_1^3
-\frac{8 \big(8 N^3+13 N^2+27 N+16\big)}{27 (N-1) N (N+1)^2} [S_1^2 + S_2]
\nonumber\\ && 
+
\Biggl[
 \frac{16 \big(35 N^4+97 N^3+178 N^2+180 N+70\big)}{27 (N-1) N (N+1)^3}
+\frac{8}{9} \bar{p}_{gq} S_2
\Biggr] S_1
\nonumber\\ && 
-\frac{16 \big(1138 N^5+4237 N^4+8861 N^3+11668 N^2+8236 N+2276\big)}{243 (N-1) N (N+1)^4}
+\frac{16}{27} \bar{p}_{gq}  S_3
\Biggr]
\nonumber
\end{eqnarray}
\begin{eqnarray}
&&
-2 \Bigg[
 3 \Biggl[
 \frac{16 \big(39 N^4+101 N^3+201 N^2+205 N+78\big)}{81 (N-1) N (N+1)^3}
+\frac{16}{27} \bar{p}_{gq}  S_2
\Biggr] S_1
\nonumber\\ &&
- \frac{16 \big(8 N^3+13 N^2+27 N+16\big)}{27 (N-1) N (N+1)^2} [S_1^2 + S_2]
+  \frac{16}{27} \bar{p}_{gq} [ S_1^3 + 2 S_3 ]
\nonumber\\ && 
-\frac{8 \big(1129 N^5+3814 N^4+8618 N^3+11884 N^2+8425 N+2258\big)}{243 (N-1) N (N+1)^4}
\Biggr]
\nonumber\\
&&
-2(2 + \textcolor{blue}{N_F}) \Biggl[
  \frac{8}{3} \bar{p}_{gq} S_1
- \frac{8 \big(8 N^3+13 N^2+27 N+16\big)}{9 (N-1) N (N+1)^2}
\Biggr] \zeta_2
+ \bar{p}_{gq} \Biggl[
 \frac{512}{9}  
-\frac{224}{9} \textcolor{blue}{N_F}
\Biggr]
\zeta_3
\Biggr\} 
\nonumber\\ && 
+ \textcolor{blue}{C_A C_F T_F}
\Biggl\{ \bar{p}_{gq} \left(96 \zeta_4 - \frac{32}{3} B_4\right)
-2 \Biggl[
 \frac{29}{27} \bar{p}_{gq} S_1^4
-\frac{2 P_8}{81 (N-1)^2 N^2 (N+1)^2 (N+2)} S_1^3
\nonumber\\ && 
+\Biggl[
\frac{2 P_9}{81 (N-1)^2 N^3 (N+1)^3 (N+2)^2}
+\frac{58}{9} \bar{p}_{gq}  S_2
\Biggr] S_1^2
+\Biggl[
\frac{32}{9} \bar{p}_{gq}  \left[\frac{53}{12}S_3 \right.
\nonumber\\ && \left.
+ S_{2,1} \right]
-\frac{4 P_{10}}{243 (N-1)^2 N^4 (N+1)^4 (N+2)^3}
-\frac{2 P_{11}}{27 (N-1)^2 N^2 (N+1)^2 (N+2)} S_2
\nonumber\\ && 
- 16 \bar{p}_{gq} S_{-2,1}
\Biggr] S_1
+\frac{2 P_{12}}{243 (N-2) (N-1)^2 N^5 (N+1)^5 (N+2)^4}
+\frac{61}{9} \bar{p}_{gq}  S_2^2
+\frac{16}{9} \bar{p}_{gq}  S_{-2}^2
\nonumber\\
&&
+\Biggl[
\frac{152}{9} \bar{p}_{gq}  S_1
-\frac{8 P_{13}}{27 (N-1)^2 N^2 (N+1)^2 (N+2)}
\Biggr] S_{-3} 
+ \frac{32 \bar{p}_{gq}}{(N-1) N (N+1) (N+2)} S_{-1} S_2
\nonumber\\ &&
+\frac{2 P_{14}}{27 (N-2) (N-1)^2 N^3 (N+1)^3 (N+2)^2} S_2
-\frac{8 P_{15}}{81 (N-1)^2 N^2 (N+1)^2 (N+2)} S_3
\nonumber\\ && 
+\frac{178}{9} \bar{p}_{gq}  S_4
+\Biggl[
 \frac{88}{9} \bar{p}_{gq} [S_1^2 + S_2]
-\frac{16 \big(52 N^4+95 N^3+210 N^2+137 N+36\big)}{27 (N-1) N^2 (N+1)^2} S_1
\nonumber\\ && 
	-\frac{32 \bar{p}_{gq} }{(N-1) N (N+1) (N+2)} S_{-1}
+\frac{8 P_{16}}{27 (N-2) (N-1)^2 N^3 (N+1)^3 (N+2)^2}
\Biggr] S_{-2}
\nonumber\\ && 
-\frac{8 \big(14 N^5+15 N^4+4 N^3+81 N^2-10 N+88\big)}{9 (N-1)^2 N^2 (N+1)^2 (N+2)} S_{2,1}
-\frac{32 \bar{p}_{gq} }{(N-1) N (N+1) (N+2)} S_{2,-1}
\nonumber\\ &&
-\frac{16}{3}  S_{3,1}
+\frac{160}{9} \bar{p}_{gq}  S_{-4}
+\frac{16 \big(26 N^4+49 N^3+126 N^2+85 N+36\big)}{27 (N-1) N^2 (N+1)^2} S_{-2,1}
-\frac{112}{9} \bar{p}_{gq} S_{-2,2}
\nonumber\\ && 
+\frac{32 \bar{p}_{gq} }{(N-1) N (N+1) (N+2)} S_{-2,-1}
-\frac{136}{9} S_{-3,1}
- 8 \bar{p}_{gq} S_{2,1,1}
+\frac{176}{9} \bar{p}_{gq} S_{-2,1,1}
\Biggr]
\nonumber\\ &&
-\frac{16}{9} \frac{(-2-2 N - N^2 +N^3)}{(N-1)N(N+1)} \left[17 S_{-3,1} + 6 S_{3,1} \right]\nonumber\\ &&
-2 
\Biggl[  \bar{p}_{gq}
\left( \frac{10}{3} S_1^2 + 2 S_2 + 4 S_{-2} \right)
-\frac{2 \big(59 N^5+94 N^4+59 N^3-84 N^2-224 N+168\big)}{9 (N-1)^2 N^2 (N+1) (N+2)} S_1
\nonumber\\ &&
+\frac{2 P_{17}}{9 (N-1) N^3 (N+1)^3 (N+2)^2} 
\Biggr]
\zeta_2
-2 
\Biggl[
\frac{2 P_{18}}{9 (N-1)^2 N^2 (N+1)^2 (N+2)}
+\frac{56}{9}  \bar{p}_{gq}  S_1
\Biggr] 
\zeta_3
\Biggr\}
\Biggr\},
\nonumber\\
\end{eqnarray}
with the polynomials $P_i$ given by
\begin{eqnarray}
P_0 &=& 205 N^8+268 N^7-3541 N^6-11258 N^5-19844 N^4-19936 N^3-9870 N^2
\nonumber\\ &&
-2016 N-216
\\
P_1 &=& 916 N^6+2547 N^5+5666 N^4+6683 N^3+4028 N^2+1200 N+144
\\
P_2 &=& 17223 N^{16}+165192 N^{15}+598459 N^{14}+791304 N^{13}-970086 N^{12}-5283698 N^{11}
\nonumber\\ &&
             -8410804 N^{10}-5109721 N^9+3632942 N^8+8912043 N^7+3490278 N^6-6793396 N^5
\nonumber\\ &&
-9096296 N^4
             -3512432 N^3+179232 N^2+131904 N+24192
\\
P_3 &=& 2708 N^{12}+12913 N^{11}+15444 N^{10}-20600 N^9-94758 N^8-142761 N^7-38594 N^6
\nonumber\\ &&
             +131744 N^5+139824 N^4+30368 N^3-5184 N-10368
\\
P_4 &=& 613 N^8+2350 N^7+3726 N^6+3208 N^5-1179 N^4-4742 N^3-2080 N^2+264 N
\nonumber\\ &&
+1296
\\
P_5 &=& N^8+3 N^7+3 N^6+3 N^5-2 N^3-16 N^2+120 N+64
\\
P_6 &=& 195 N^9+1152 N^8+2431 N^7+1949 N^6-1038 N^5-4577 N^4-4568 N^3-2376 N^2
\nonumber\\ &&
-320 N
-912
\\
P_7 &=& 233 N^8+980 N^7+1492 N^6+926 N^5+39 N^4-426 N^3+1068 N^2+1352 N+480
\\
P_8 &=& 371 N^6+945 N^5+945 N^4-N^3-2132 N^2-728 N+744
\\
P_9 &=& 1540 N^9+8277 N^8+18725 N^7+20707 N^6-1501 N^5-31700 N^4-20384 N^3
\nonumber\\ &&
+9664 N^2
+7056 N-1152
\\
P_{10} &=&
5042 N^{12}+42402 N^{11}+155879 N^{10}+312989 N^9+312249 N^8-24387 N^7
\nonumber\\ &&
-415670 N^6
-329640 N^5+53392 N^4
+118912 N^3-17472 N^2-33984 N
-3456
\\
P_{11} &=& 319 N^6+765 N^5+945 N^4+523 N^3-1696 N^2-592 N-120
\\
P_{12} &=& 10393 N^{16}+100762 N^{15}+401681 N^{14}+757324 N^{13}+161109 N^{12}
          -2974044 N^{11}
\nonumber\\ &&
-8597809 N^{10}-11649324 N^9-5136842 N^8+7823954 N^7
          +15364644 N^6
+13824600 N^5
\nonumber\\ &&
+8495280 N^4+3677408 N^3+943296 N^2+153216 N+34560
\\
P_{13} &=& 52 N^6+159 N^5+291 N^4+253 N^3-373 N^2-334 N-192
\\
P_{14} &=& 324 N^{10}+975 N^9+347 N^8-2523 N^7-7715 N^6-7672 N^5+8128 N^4+14168 N^3
\nonumber\\ &&
           +2480 N^2 +4448 N+4608
\\
P_{15} &=& 184 N^6+432 N^5+675 N^4+484 N^3-1087 N^2-268 N+84
\\
P_{16} &=& 136 N^{10}+503 N^9+285 N^8-1445 N^7-4499 N^6-5032 N^5+1254 N^4+5838 N^3
\nonumber\\ &&
+3640 N^2
-1304 N -960
\\
P_{17} &=& 40 N^8+247 N^7+787 N^6+1771 N^5+2775 N^4+2564 N^3+1384 N^2+704 N+240
\\
P_{18} &=& 41 N^6+315 N^5+139 N^4-215 N^3+1028 N^2-236 N+1136~.
\end{eqnarray}
$a_{qg}^{(3)}(N)$ contains contributions up to weight {\sf w = 4}, including the constant
\begin{eqnarray}
B_4 &=&-4\zeta_2\ln^2(2) +\frac{2}{3}\ln^4(2)
-\frac{13}{2}\zeta_4 +16 {\rm Li}_4\Bigl(\frac{1}{2}\Bigr) = -8 S_{-3,-1}(\infty) + 
\frac{11}{2} \zeta_4.
\label{eqB4} 
\end{eqnarray}
After algebraic reduction  \cite{Blumlein:2003gb} the result is represented in terms 
of the basis~:
\begin{eqnarray}
S_1, S_2, S_3, S_4, S_{-1}, S_{-2}, S_{-3}, S_{-4}, S_{2,1}, S_{2,-1}, S_{-2,-1}, S_{-2,1},
S_{-2,2}, S_{3,1}, S_{-3,1}, S_{2,1,1}, S_{-2,1,1}~.
\end{eqnarray}
Due to structural relations, such as differentiation and multiple argument relations, \cite{Blumlein:2009ta}, 
only the representatives    
\begin{eqnarray}
S_1, S_{2,1}, S_{-2,1}, S_{-3,1}, S_{2,1,1}, S_{-2,1,1}
\end{eqnarray}
remain as basic sums. The function $a_{gq}^{(3)}(N)$ contains a removable singularity at $N=2$ 
yielding
\begin{eqnarray}
a_{gq}^{(3)}(N \rightarrow 2) &=&
- C_F C_A T_F \left[\frac{414817}{2187}   + \frac{296}{9} \zeta_3 \right]
- C_F^2 T_F   \left[\frac{1098203}{2187}  + 16 \zeta_3 \right]
\end{eqnarray}
for the evanescent pole term. This is in accordance with the expectation, that the gluonic OMEs have their rightmost 
singularity at $N=1$. In massive OMEs, removable singularities can in general also occur at rational values of $N > 
1$, 
cf.~\cite{Blumlein:2013ota}. $A_{gq}(N)$ can be represented by harmonic sums \cite{HSUM} over $\mathbb{Q}(N)$ 
with rational weights whose denominators factorize into terms 
$(N-k)^l, k \in \mathbb{Z}, l \in \mathbb{N}$. It is therefore a meromorphic function \cite{Blumlein:2009ta}. 
More specifically, its poles are located at the integers $N \leq 1$. 

Let us finally consider the behaviour of $a_{qg}^{(3)}(N)$ for $N \rightarrow \infty$ and around the so-called
`leading singularity' at $N = 1$. 

For large values of $N \in \mathbb{C}$ outside the singularities one obtains
\begin{eqnarray}
a_{qg}^{(3)}(N \rightarrow \infty) &\propto& 
- \frac{58}{27 N} (C_A-C_F) C_F T_F {L}^4(N)
\nonumber\\ &&
-\frac{2 C_F T_F}{81 N} \left[-742 C_A + 550 C_F + 24 T_F(N_F + 2)\right] {L}^3(N) 
\nonumber\\ &&
+ \frac{C_F T_F}{N} \left[
  C_A \left(-\frac{88}{9} \zeta_2 -\frac{6160}{81}\right)
+ C_F \left(\frac{32}{9} \zeta_2 +\frac{3664}{81}\right) \right.
\nonumber\\ && \left.
+ \frac{128}{27} (N_F+2) T_F\right] {L}^2(N) 
+ O\left(\frac{{L}(N)}{N}\right)
\end{eqnarray}
as the first terms in the asymptotic representation, with ${L}(N) = \ln(N) + \gamma_E$ and $\gamma_E$
denotes the Euler-Mascheroni constant. In $x$-space the leading singular term is $\propto \alpha_s^3 
\ln^4(1-x)$ for $x \rightarrow 1$.

Analyzing the anomalous dimensions for different scattering processes in fixed order perturbation theory one
finds that so-called `leading-poles' are situated at $N = 1$ for massless vector operators \cite{Gross:1974cs}, 
$N = 0$ in case of massless quark operators \cite{Kirschner:1983di,Blumlein:1995jp}, and $N = -1$ for massless 
scalar operators \cite{Blumlein:1998mg}. Expanding $a_{qg}^{(3)}(N)$ around $N = 1$ the leading term  is given by
\begin{eqnarray}
a_{qg}^{(3)}(N \rightarrow 1) \hspace*{-1mm} &\propto& \hspace*{-2mm} \frac{1}{(N-1)^2}\Biggl[
C_A C_F T_F \Biggl(
\frac{16}{3} \zeta_2 -\frac{736}{9} \zeta_3 + \frac{20224}{81}\Biggr)
+ C_F^2 T_F \Biggl(\frac{224}{9} \zeta_2 +\frac{1024}{9} \zeta_3 
\nonumber\\ && -\frac{37376}{243}
\Biggr) \Biggr]
+O\left(\frac{1}{N-1}\right)~.
\end{eqnarray}
In QCD, $C_A = 3, C_F = 4/3, T_F = 1/2$, the first expansion coefficients are given by
\begin{eqnarray}
a_{qg}^{(3)}(N) &\propto& 
\frac{341.543}{(N-1)^2}
-\frac{1}{N-1}\left(1814.73
-66.0055 N_F\right)
+3222.81 
-71.4816 N_F
\nonumber\\ &&
-\left(5345.61 
- 81.4031 N_F \right) (N-1)
+\left(8454.89
-85.7885 N_F \right) (N-1)^2 
\nonumber\\ &&
+ O\left((N-1)^3\right)~.
\end{eqnarray}
In $x$-space the leading term is of $\propto \alpha_s^3 \ln(1/x)/x$. The values of the sub-leading
terms have oscillating signs and rise from term to term, which leads to a strong compensation
of the leading behaviour in the physical region, e.g. at HERA. This is in accordance with earlier 
observations in other cases, cf. 
Refs.~\cite{Blumlein:1995jp,Blumlein:1996hb,Blumlein:1997em,Blumlein:1998mg}.
The complete OME $A_{gq,Q}^{(2,3), \MS}$ in $N$ and $x$-space are given in the Appendix.
\section{Conclusions}
\label{sec:5}

\vspace*{1mm}
\noindent
We calculated the massive operator matrix element $A_{gq}^{(3)}(N)$, which represents the first
complete transition matrix element in the variable flavor scheme at 3-loop order. The corresponding
Feynman integrals have been reduced using the integration-by-parts technique to master integrals,
which have been computed using different techniques in terms of generating functions. In Mellin-space
the final result for the individual diagrams has been obtained using difference-field techniques.
The matrix element $A_{gq}^{(3)}(N)$ can be expressed by harmonic sums up to weight {\sf w = 4} in Mellin 
space and harmonic polylogarithms up to weight {\sf w = 5} in $x$-space. This simple form may be
due to the fact that at most four of the lines of the corresponding diagrams are massive. The contributing
graphs include diagrams of the Benz-topology. Both the results for renormalizing the heavy quark mass in the 
on-shell and $\overline{\rm MS}$ scheme were presented. As a by-product of the calculation also the contribution 
$\propto T_F$ to the 3-loop anomalous dimension $\gamma_{gq}(N)$ has been obtained and confirms the results in the 
literature for the first time. 
 
\section{Appendix}
\label{sec:A}

\vspace*{1mm}
\noindent
In the $\MS$-scheme for the strong coupling and the on-shell-scheme of the heavy quark mass 
$m$ the massive OME $A_{gq,Q}(N)$ up to 3-loop order is given in Mellin space by~:
\begin{eqnarray}
A_{gq,Q}^{\rm OMS}(N, a_s) &=&    a_s^2 A_{gq,Q}^{(2),\rm OMS}(N)
                                + a_s^3 A_{gq,Q}^{(3),\rm OMS}(N)
\end{eqnarray}
with $a_s = \alpha_s^{\MS}(\mu^2)/(4\pi)$ and
\begin{eqnarray}
A_{gq,Q}^{(2),\rm OMS}(N) &=& 
 \frac{1}{2}\left[1+(-1)^N\right]
\textcolor{blue}{C_F T_F}
\Biggl\{
\frac{8}{3} \pgq \textcolor{blue}{\ln^2\left(\frac{m^2}{\mu^2}\right)}
+\Biggl[
\frac{16 \big(8 N^3+13 N^2+27 N+16\big)}{9 (N-1) N (N+1)^2}
\nonumber\\ &&
-\frac{16}{3} \pgq S_1\Biggr] 
\textcolor{blue}{\ln\left(\frac{m^2}{\mu^2}\right)}
+\frac{4}{3} \pgq  \left[S_1^2 + S_2\right]
-\frac{8 \big(8 N^3+13 N^2+27 N+16\big)}{9 (N-1) N (N+1)^2} S_1
\nonumber\\ &&
+\frac{8 \big(43 N^4+105 N^3+224 N^2+230 N+86\big)}{27 (N-1) N (N+1)^3}
\Biggr\}
\end{eqnarray}
\begin{eqnarray}
A_{gq,Q}^{(3),\rm OMS}(N) &=& \frac{1}{2}\left[1+(-1)^N\right] \nonumber\\ &&
\times
\Biggl\{
\pgq 
\Biggl\{
\textcolor{blue}{C_F^2 T_F} 
\Biggl[
\frac{4 P_{32}}{9 (N-1) N^2 (N+1)^2 (N+2)}
-\frac{16}{9} S_1
\Biggr]
+ \frac{32}{9}
\textcolor{blue}{C_F T_F^2}
(2 + \textcolor{blue}{N_F})
\nonumber\\ &&
+\textcolor{blue}{C_F C_A T_F} 
\Biggl[
\frac{16}{9} S_1
-\frac{8 P_{20}}{9 (N-1) N (N+1) (N+2)}
\Biggr]
\Biggr\} 
\textcolor{blue}{\ln^3\left(\frac{m^2}{\mu^2}\right)}
\nonumber\\ &&
+
\Biggl\{
\textcolor{blue}{C_F^2 T_F} 
\Biggl\{
-\frac{2 P_{47}}{9 (N-1)^2 N^3 (N+1)^3 (N+2)^2}
+\frac{16 \big(7 N^3+4 N^2+17 N-6\big)}{9 (N-1) N^2 (N+1)} S_1
\nonumber\\  && 
+
\frac{8}{3} (S_1^2 - 5 S_2)
\pgq 
\Biggr\} 
+ \textcolor{blue}{C_F T_F^2}
\Biggl\{
\frac{32 \big(8 N^3+13 N^2+27 N+16\big)}{9 (N-1) N (N+1)^2}
-\frac{32}{3} \pgq S_1
\Biggr\}
\nonumber\\ && 
+ \textcolor{blue}{C_F C_A T_F}
\Biggl\{
\frac{4 P_{46}}{9 (N-1)^2 N^3 (N+1)^3 (N+2)^2}
\nonumber\\ && 
+\frac{32 P_{26}}{9 (N-1)^2 N^2 (N+1) (N+2)} S_1
+ 
\Biggl[
-\frac{8}{3} S_1^2-8 S_2
-16 S_{-2}
\Biggr] 
\pgq 
\Biggr\}\Biggr\}
\textcolor{blue}{\ln^2\left(\frac{m^2}{\mu^2}\right)}
\nonumber\\ &&
+
\Biggl\{
\textcolor{blue}{C_F^2 T_F}
\Biggl\{
-\frac{4 P_{23}}{9 (N-1) N^2 (N+1)^2} S_1^2
+\frac{8 P_{39}}{27 (N-1) N^3 (N+1)^3} S_1
\nonumber\\ &&
-\frac{2 P_{54}}{27 (N-1)^2 N^5 (N+1)^4 (N+2)^2}
-\frac{4 P_{44}}{9 (N-1)^2 N^3 (N+1)^3 (N+2)} S_2
\nonumber\\ && 
+
\Biggl[
-\frac{8}{9} S_1^3
+\frac{88}{3} S_2 S_1
+\frac{176}{9} S_3
-\frac{128}{(N-1) N (N+1) (N+2)} S_{-2}
-\frac{32}{3} S_{2,1}
\nonumber\\ && 
-64 \zeta_3
\Biggr] 
\pgq 
\Biggr\}
+ \textcolor{blue}{C_F T_F^2 N_F}
\Biggl\{
\frac{32 \big(8 N^3+13 N^2+27 N+16\big)}{9 (N-1) N (N+1)^2} S_1
\nonumber\\ && 
+\frac{32 P_{21}}{27 (N-1) N (N+1)^3}
+
\Biggl[
-\frac{16}{3} S_1^2
-\frac{16}{3} S_2
\Biggr] \pgq
\Biggr\} + \textcolor{blue}{C_F T_F^2} \frac{992}{27} \pgq
\nonumber\\ && 
+
\textcolor{blue}{C_F C_A T_F}
\Biggl\{
\frac{4 P_{29}}{9 (N-1)^2 N^2 (N+1) (N+2)} S_1^2
\nonumber\\ &&
-\frac{8 P_{49}}{27 (N-1)^2 N^3 (N+1)^3 (N+2)^2} S_1
+\frac{8 P_{53}}{27 (N-1)^2 N^4 (N+1)^4 (N+2)^3}
\nonumber\\ &&
-\frac{4 P_{35}}{3 (N-1)^2 N^2 (N+1)^2 (N+2)} S_2
-\frac{16 P_{34}}{3 (N-1)^2 N^2 (N+1)^2 (N+2)} S_{-2}
\nonumber\\ && 
+ \Biggl[
\frac{8}{9} S_1^3
+\frac{56}{3} S_2 S_1
+\frac{128}{3} S_{-2} S_1
+\frac{256}{9} S_3
+\frac{64}{3} S_{-3}
-\frac{64}{3} S_{-2,1}
\nonumber\\ && 
+64 \zeta_3
\Biggr] \pgq 
\Biggr\}\Biggr\}
\textcolor{blue}{\ln\left(\frac{m^2}{\mu^2}\right)}
+ \textcolor{blue}{C_F^2 T_F}
\Biggr\{
\frac{40 \big(4 N^3+N^2+11 N-6\big)}{81 (N-1) N^2 (N+1)} S_1^3
\nonumber\\ && 
-\frac{8 P_{33}}{81 (N-1) N^3 (N+1)^3} S_1^2
+\Biggl[
\frac{16 P_{45}}{243 (N-1) N^4 (N+1)^4}
\nonumber\\ &&
+\frac{8 P_{25}}{27 (N-1) N^2 (N+1)^2} S_2
\Biggr]
S_1
+\frac{8 P_{43}}{9 (N-1)^2 N^3 (N+1)^3 (N+2)} \zeta_3
\nonumber\\ &&
+\frac{P_{57}}{486 (N-2) (N-1)^2 N^5 (N+1)^6 (N+2)^4}
\nonumber\\ &&
-\frac{8 P_{52}}{81 (N-2) (N-1)^2 N^4 (N+1)^4 (N+2)} S_2
\nonumber\\ &&
+\frac{16 P_{41}}{81 (N-1)^2 N^2 (N+1)^2 (N+2)} S_3
\nonumber\\ &&
-\frac{32 P_{42}}{3 (N-2) (N-1)^2 N^3 (N+1)^3 (N+2)^2} S_{-2}
\nonumber\\ &&  
-\frac{16 \big(35 N^3+64 N^2+111 N+70\big)}{27 (N-1) N (N+1)^2} S_{2,1}
+
\Biggl[
\frac{10}{27} S_1^4-\frac{76}{9} S_2 S_1^2
+\Biggl[
-\frac{304}{27} S_3
+\frac{32}{9} S_{2,1}
\nonumber\\ && 
+\frac{320}{9} \zeta_3
\Biggr] 
S_1
-\frac{2}{3} S_2^2
- 96 \zeta_4 
+\frac{64}{3} B_4
+\frac{128}{(N-1) N (N+1) (N+2)} S_{-1} S_2
-\frac{172}{9} S_4
\nonumber\\ && 
-\frac{128}{(N-1) N (N+1) (N+2)} S_{-2} S_{-1}
+\frac{128}{3 (N-1) N (N+1) (N+2)} S_{-3}
\nonumber\\ && 
-\frac{128}{(N-1) N (N+1) (N+2)} S_{2,-1}
+\frac{32}{3} S_{3,1}
+\frac{128}{(N-1) N (N+1) (N+2)} S_{-2,-1}
\nonumber\\ && 
-\frac{32}{9} S_{2,1,1}
\Biggr] 
\pgq 
\Biggr\}
+
\textcolor{blue}{C_F T_F^2} 
\Biggl\{
\Biggl[
-\frac{16 \big(8 N^3+13 N^2+27 N+16\big)}{27 (N-1) N (N+1)^2} S_1^2
\nonumber\\ && 
+\frac{32 P_{19}}{27 (N-1) N (N+1)^3} S_1
-\frac{16 P_{31}}{243 (N-1) N (N+1)^4}
\nonumber\\ &&
-\frac{16 \big(8 N^3+13 N^2+27 N+16\big)}{27 (N-1) N (N+1)^2} S_2
+\textcolor{blue}{N_F} 
\Biggl[
-\frac{32 \big(8 N^3+13 N^2+27 N+16\big)}{27 (N-1) N (N+1)^2} S_1^2
\nonumber\\ &&
+\frac{64 P_{19}}{27 (N-1) N (N+1)^3} S_1
+\frac{64 P_{30}}{243 (N-1) N (N+1)^4}
\nonumber\\ && 
-\frac{32 \big(8 N^3+13 N^2+27 N+16\big)}{27 (N-1) N (N+1)^2} S_2
\Biggr]
+
\Biggl[
\frac{16}{27} S_1^3
+\frac{16}{9} S_2 S_1
+\frac{32}{27} S_3
\nonumber\\ && 
+
\textcolor{blue}{N_F} 
\Biggl[
\frac{32}{27} S_1^3
+\frac{32}{9} S_2 S_1
+\frac{64}{27} S_3
-\frac{256}{9} \zeta_3
\Biggr]
+\frac{448}{9} \zeta_3
\Biggr] \pgq
\Biggr\}
\nonumber\\ && 
+ \textcolor{blue}{C_F C_A T_F}
\Biggl\{
-\frac{16 P_{27}}{81 (N-1)^2 N^2 (N+1) (N+2)} S_1^3
\nonumber\\ &&
+\frac{4 P_{48}}{81 (N-1)^2 N^3 (N+1)^3 (N+2)^2} S_1^2
+\Biggl[
\frac{16 P_{37}}{27 (N-1)^2 N^2 (N+1)^2 (N+2)} S_2
\nonumber\\ &&
-\frac{8 P_{55}}{243 (N-1)^2 N^4 (N+1)^4 (N+2)^3}
\Biggr] 
S_1
-\frac{4 P_{36}}{9 (N-1)^2 N^2 (N+1)^2 (N+2)} \zeta_3
\nonumber\\ &&
+\frac{8 P_{56}}{243 (N-2) (N-1)^2 N^5 (N+1)^5 (N+2)^4}
+
\Biggl[
\frac{32 P_{24}}{27 (N-1) N^2 (N+1)^2} S_1
\nonumber\\ &&
-\frac{16 P_{51}}{27 (N-2) (N-1)^2 N^3 (N+1)^3 (N+2)^2}
\Biggr] 
S_{-2}
\nonumber\\ &&
-\frac{4 P_{50}}{27 (N-2) (N-1)^2 N^3 (N+1)^3 (N+2)^2} S_2
\nonumber\\ &&
+\frac{8 P_{40}}{81 (N-1)^2 N^2 (N+1)^2 (N+2)} S_3
+\frac{16 P_{38}}{27 (N-1)^2 N^2 (N+1)^2 (N+2)} S_{-3}
\nonumber\\ &&
+\frac{16 P_{28}}{9 (N-1)^2 N^2 (N+1)^2 (N+2)} S_{2,1}
-\frac{32 P_{22}}{27 (N-1) N^2 (N+1)^2} S_{-2,1}
\nonumber\\ && 
+
\pgq 
\Biggl[
-\frac{10}{27} S_1^4
-\frac{68}{9} S_2 S_1^2
-\frac{304}{9} S_{-3} S_1
+\Biggl[
-\frac{752}{27} S_3
-\frac{64}{9} S_{2,1}
+32 S_{-2,1}
\nonumber\\ && 
-\frac{128}{9} \zeta_3
\Biggr] 
S_1
-\frac{122}{9} S_2^2
-\frac{32}{9} S_{-2}^2
+ 96 \zeta_4
-\frac{64}{(N-1) N (N+1) (N+2)} S_{-1} S_2
\nonumber\\ && 
-\frac{356}{9} S_4
+
\Biggl[
-\frac{176}{9} S_1^2
-\frac{176}{9} S_2
+\frac{64}{(N-1) N (N+1) (N+2)} S_{-1}
\Biggr] 
S_{-2}
\nonumber\\ && 
-\frac{320}{9} S_{-4}
+\frac{64}{(N-1) N (N+1) (N+2)} S_{2,-1}
+\frac{32}{3} S_{3,1}
+\frac{224}{9} S_{-2,2}
\nonumber\\ && 
-\frac{64}{(N-1) N (N+1) (N+2)} S_{-2,-1}
+\frac{272}{9} S_{-3,1}
+16 S_{2,1,1}
-\frac{352}{9} S_{-2,1,1}
\nonumber\\ &&
-\frac{32}{3} B_4 
\Biggr]
\Biggr\}
\Biggr\},
\end{eqnarray}
with the polynomials $P_i$
\begin{eqnarray}
P_{19} &=& 4 N^4+4 N^3+23 N^2+25 N+8
\\
P_{20} &=& 11 N^4+22 N^3-7 N^2-18 N+4
\\
P_{21} &=& 19 N^4+81 N^3+86 N^2+80 N+38
\\
P_{22} &=& 26 N^4+49 N^3+126 N^2+85 N+36
\\
P_{23} &=& 43 N^4+68 N^3+135 N^2+74 N-24
\\
P_{24} &=& 52 N^4+95 N^3+210 N^2+137 N+36
\\
P_{25} &=& 166 N^4+293 N^3+546 N^2+341 N+18
\\
P_{26} &=& N^5-N^4-8 N^3-33 N^2-7 N+30
\\
P_{27} &=& 4 N^5+23 N^4+49 N^3+165 N^2-N-150
\\
P_{28} &=& 14 N^5+15 N^4+4 N^3+81 N^2-10 N+88
\\
P_{29} &=& 31 N^5+86 N^4+127 N^3+324 N^2-88 N-264
\\
P_{30} &=& 197 N^5+824 N^4+1540 N^3+1961 N^2+1388 N+394
\\
P_{31} &=& 359 N^5+1364 N^4+2944 N^3+3608 N^2+2495 N+718
\\
P_{32} &=& 3 N^6+9 N^5+17 N^4+19 N^3+52 N^2+44 N+48
\\
P_{33} &=& 4 N^6-216 N^5-343 N^4-694 N^3-847 N^2-456 N-72
\\
P_{34} &=& 13 N^6+37 N^5+51 N^4+39 N^3-76 N^2-72 N-40
\\
P_{35} &=& 15 N^6+33 N^5+37 N^4-N^3-92 N^2+56
\\
P_{36} &=& 19 N^6+249 N^5+65 N^4-253 N^3+1084 N^2-172 N+1120
\\
P_{37} &=& 31 N^6+54 N^5+72 N^4+19 N^3-283 N^2-7 N-66
\\
P_{38} &=& 52 N^6+159 N^5+291 N^4+253 N^3-373 N^2-334 N
\nonumber\\ &&
-192
\\
P_{39} &=& 164 N^6+243 N^5+472 N^4+175 N^3-446 N^2-408 N-72
\\
P_{40} &=& 269 N^6+567 N^5+981 N^4+725 N^3-2066 N^2-356 N+24
\\
P_{41} &=& 275 N^6+774 N^5+828 N^4+434 N^3-1631 N^2-1532 N-876
\\
P_{42} &=& N^8+3 N^7+3 N^6+3 N^5-2 N^3-16 N^2+120 N+64
\\
P_{43} &=& 115 N^8+484 N^7+730 N^6+436 N^5-33 N^4-280 N^3+436 N^2+608 N+192
\\
P_{44} &=& 187 N^8+706 N^7+1056 N^6+802 N^5-807 N^4-2036 N^3-1444 N^2-480 N
\nonumber\\ &&
-288
\\
P_{45} &=& 1031 N^8+3236 N^7+5986 N^6+5051 N^5-1186 N^4-6041 N^3-4161 N^2
\nonumber\\ &&
-1008 N-108
\\
P_{46} &=& 69 N^9+414 N^8+860 N^7+694 N^6-309 N^5-820 N^4-124 N^3+528 N^2
\nonumber\\ &&
+896 N+384
\\
P_{47} &=& 87 N^9+450 N^8+604 N^7-526 N^6-2451 N^5-3236 N^4-1616 N^3-720 N^2
\nonumber\\ &&
+112 N-768
\\
P_{48} &=& 284 N^9+1821 N^8+5119 N^7+12689 N^6+22753 N^5+16190 N^4-7888 N^3
\nonumber\\ &&
-17128 N^2-4752 N+1152
\\
P_{49} &=& 356 N^9+2139 N^8+5347 N^7+9317 N^6+11533 N^5+4460 N^4-7168 N^3
\nonumber\\ &&
-8464 N^2-1680 N+576
\\
P_{50} &=& 78 N^{10}+81 N^9-265 N^8-759 N^7-1853 N^6+434 N^5+10204 N^4+10352 N^3
\nonumber\\ &&
+2528 N^2+5984 N+4608
\\
P_{51} &=& 136 N^{10}+503 N^9+285 N^8-1445 N^7-4499 N^6-5032 N^5+1254 N^4+5838 N^3
\nonumber\\ &&
+3640 N^2-1304 N-960
\\
P_{52} &=& 1132 N^{11}+3042 N^{10}+297 N^9-8170 N^8-20968 N^7-14788 N^6+14047 N^5
\nonumber\\ &&
+21932 N^4-6172 N^3
           -8496 N^2-7776 N-5184
\\
P_{53} &=& 301 N^{12}+2301 N^{11}+7232 N^{10}+13239 N^9+14891 N^8+8033 N^7+5814 N^6
\nonumber\\ &&
+23063 
N^5+44966 N^4
           +49136 N^3+28928 N^2+7440 N+288
\\
P_{54} &=& 339 N^{12}+1545 N^{11}+436 N^{10}-11406 N^9-31663 N^8-32981 N^7+3704 N^6
\nonumber\\ &&
+ 44386 N^5+81160 N^4
           +65344 N^3+18240 N^2+9504 N+3456
\\
P_{55} &=& 1207 N^{12}+10131 N^{11}+39379 N^{10}+120121 N^9+299724 N^8+482244 N^7
\nonumber\\ &&
+382214 N^6-15780 N^5
           -256840 N^4-118240 N^3+39648 N^2
\nonumber\\ &&
+33984 N+3456
\\
P_{56} &=& 2596 N^{16}+23287 N^{15}+76658 N^{14}+88963 N^{13}-121050 N^{12}-504489 N^{11}
\nonumber\\ &&
-386836 N^{10}
           +625443 N^9+1086304 N^8-574564 N^7-3300816 N^6-4766232 N^5
\nonumber\\ &&
-3890160 N^4-1844224 N^3-441984 N^2
           -47232 N-6912
\\
P_{57} &=& -138495 N^{17}-1469301 N^{16}-6177407 N^{15}-11396201 N^{14}+794307 N^{13}
\nonumber\\ &&
+49867573 N^{12}
           +113250363 N^{11}+119019157 N^{10}+23906680 N^9-106346044 N^8
\nonumber\\ &&
-139149336 N^7-31178992 N^6
           +90199968 N^5+96114752 N^4+36681856 N^3
\nonumber\\ &&
+7448064 N^2+3036672 N+718848~.
\end{eqnarray}
The analytic continuation of the OMEs $A_{gq,Q}^{\rm S, (2,3)}(N)$ to complex values of $N$
are obtained as has been described in 
Refs.~\cite{Blumlein:2000hw,Blumlein:2005jg,Blumlein:2009ta,Blumlein:2009fz}.

For the $x$-space representation it is convenient to define 
\begin{eqnarray}
p_{gq}(x) = \frac{1}{x}\left[1+(1-x)^2\right]~.
\end{eqnarray}
The operator matrix elements $A_{gq,Q}^{(2,3), \rm OMS}(x)$ can be expressed in terms of harmonic
polylogarithms \cite{Remiddi:1999ew}, for which we use the shorthand notation 
$H_{\vec{a}}(x) \equiv H_{\vec{a}}$. It reads~:
\begin{eqnarray}
A_{gq,Q}^{(2), \rm OMS}(x) &=& \textcolor{blue}{C_F T_F} 
\Biggl\{
\frac{8}{3} p_{gq} \textcolor{blue}{\ln^2\left(\frac{m^2}{\mu^2}\right)}
+ \Biggl[
\frac{32 \big(4 x^2-5 x+5\big)}{9 x}
-\frac{16}{3} p_{gq} H_1
\Biggr] \textcolor{blue}{\ln\left(\frac{m^2}{\mu^2}\right)}
\nonumber\\ &&
+\frac{4}{3} p_{gq} H_1^2
-\frac{16 \big(4 x^2-5 x+5\big)}{9 x} H_1
+\frac{8 \big(43 x^2-56 x+56\big)}{27 x}
\Biggr\}
\\
A_{gq,Q}^{(3), \rm OMS}(x) &=& \Biggl\{
\textcolor{blue}{C_F C_A T_F} \Biggl[
 \frac{32 \big(x^2+x+1\big)}{9 x} H_0
+\frac{16}{9} p_{gq} H_1
-\frac{8 \big(4 x^3+36 x^2-42 x+35\big)}{27 x}
\Biggr]
\nonumber\\ &&
+ \textcolor{blue}{C_F T_F^2}(\textcolor{blue}{N_F} + 2) \frac{32}{9} p_{gq} 
+ \textcolor{blue}{C_F^2 T_F} \Biggl[
 \frac{8 \big(17 x^2+2 x-16\big)}{9 x} H_0
-\frac{16}{9} p_{gq} H_1
\nonumber\\ &&
-\frac{16}{3} (x-2) H_0^2
+\frac{4 \big(32 x^3-327 x^2+552 x-248\big)}{27 x}
\Biggr]
\Biggr\} \textcolor{blue}{\ln^3\left(\frac{m^2}{\mu^2}\right)}
\nonumber\\ &&
+ \Biggl\{\textcolor{blue}{C_F C_A T_F} 
\Biggl\{
p_{gq} 
\Biggl[
 \frac{16}{3} [H_0 H_1 - H_{0,1}]
-\frac{8}{3} H_1^2 
\Biggr]
+ \frac{16(x^2+2x+2)}{x} [H_0 H_{-1} - H_{0,-1}] 
\nonumber\\ &&
-\frac{16 \big(x^2+10 x-2\big)}{3 x} H_{0,1}
-\frac{16 \big(8 x^3-4 x^2+32 x+3\big)}{9 x} H_0
\nonumber\\ &&
+\frac{16 \big(4 x^3+5 x^2+14 x-21\big)}{9 x} H_1
+\frac{16}{3} (x+4) H_0^2
+\frac{32(2+x)(1 + 2 x)}{3 x} \zeta_2
\nonumber\\ && 
+\frac{4 \big(68 x^3-104 x^2+296 x-191\big)}{9 x}
\Biggr\}
+ \textcolor{blue}{C_F T_F^2}
\Biggl\{
\frac{64 \big(4 x^2-5 x+5\big)}{9 x}
-\frac{32}{3} p_{gq} H_1 \Biggr\}
\nonumber\\ &&
+ \textcolor{blue}{C_F^2 T_F}
\Biggl\{
p_{gq}  \left[
 16 \left[H_0 H_1 - H_{0,1}\right]
+\frac{8}{3} H_1^2
\right]
+\frac{16}{3} (x-2) H_{0,1}
\nonumber\\ &&
+\frac{16 \big(13 x^2-23 x+17\big)}{9 x} H_1
-\frac{4 \big(32 x^3-191 x^2-134 x+56\big)}{9 x} H_0
-\frac{8}{3} (13 x-8) H_0^2
\nonumber\\ &&
+\frac{2 \big(896 x^3-3945 x^2+3468 x-680\big)}{27 x}
-\frac{16}{3} (x-2) \zeta_2
\Biggr\} 
\Biggr\}
\textcolor{blue}{\ln^2\left(\frac{m^2}{\mu^2}\right)}
\nonumber\\ &&
+\Biggl\{
\textcolor{blue}{C_F^2 T_F}
\Biggl\{
\frac{4}{3} (x-2) H_0^4
+\frac{4}{3} (7 x-2) H_0^3
-\frac{2}{9} (973 x+640) H_0^2
\nonumber\\ &&
-\frac{32 (x+1) \big(4 x^2-31 x+31\big)}{9 x} H_{-1} H_0
+\Biggl[
-\frac{4 \big(480 x^3-3656 x^2-3407 x-528\big)}{27 x}
\nonumber\\ &&
-\frac{32 \big(5 x^2+2 x+4\big) \zeta_2}{3 x}
+64 (x-2) \zeta_3
\Biggr]
H_0
-\frac{8}{9} \ppgq H_1^3
+\Biggl[
-\frac{4 \big(67 x^2-110 x+86\big)}{9 x}
\nonumber\\ &&
-16 \ppgq H_0
\Biggr]
 H_1^2
+ 96 (x-2) \zeta_4
+\frac{2 \big(3616 x^3-15955 x^2 +7324 x+4676\big)}{27 x}
\nonumber\\ &&
+\Biggl[
\frac{32 (x+1) \big(4 x^2-31 x+31\big)}{9 x}
-\frac{64 \big(3 x^2-6 x+2\big)}{3 x} H_0
\Biggr]
H_{0,-1}
\nonumber\\ &&
+\Biggl[
\frac{16 \big(8 x^3-116 x^2+145 x-104\big)}{9 x}
-\frac{16 \big(3 x^2-4\big)}{x} H_0
+\frac{32}{3} 
\ppgq H_1
\Biggr]
H_{0,1}
\nonumber\\ &&
+\frac{128 \big(3 x^2-6 x+2\big)}
{3 x} H_{0,0,-1}
+\Biggl[
32 (x-2) H_0
-\frac{32 \big(10-9 x^2\big)}{3 x}
\Biggr]
H_{0,0,1}
+\frac{64}{3 x} H_{0,1,1}
\nonumber\\ &&
-96 (x-2) H_{0,0,0,1}
+\frac{16 \big(65 x^2+20 x-62\big)}{9 x} \zeta_2
+ \Biggl[
\frac{16 \big(115 x^2+16 x-49\big)}{27 x}
\nonumber\\ &&
-\frac{16 \big(8 x^3-105 x^2+165 x-104\big)}{9 x} H_0
-\frac{16}{3} \ppgq H_0^2
+\frac{64}{3} \ppgq \zeta_2
\Biggr]
H_1
\nonumber\\ &&
-\frac{64 \big(12 x^2-15 x+5\big)}{3 x} \zeta_3
\Biggr\}
+ \textcolor{blue}{C_F T_F^2} \frac{992}{27} \ppgq 
+\textcolor{blue}{C_F T_F^2 N_F}
\Biggl\{
-\frac{16}{3} \ppgq H_1^2
\nonumber\\ && 
+\frac{64 \big(4 x^2-5 x+5\big)}{9 x} H_1
+\frac{32 \big(19 x^2-68 x +68\big)}{27 x}
\Biggr\}
+\textcolor{blue}{C_F C_A T_F}
\Biggl\{
-\frac{8}{9} (x+2) H_0^3
\nonumber\\ && 
+\Biggl[
\frac{4}{9} \big(8 x^2+151 x+292\big)
-\frac{32 \big(x^2+2 x +2\big)}{3 x} H_{-1}
\Biggr]
H_0^2
+\Biggl[
\frac{64 \big(x^2+2 x+2\big)}{3 x} H_{-1}^2
\nonumber\\ && 
+\frac{16 \big(4 x^3-9 x^2+30 x+82\big)}{9 x} H_{-1}
-\frac{8 \big(424 x^3+136 x^2+1387 x+356\big)}{27 x}
\nonumber\\ && 
+\frac{32 \big(5 x^2+4 x+2\big)}{3 x} \zeta_2
\Biggr]
H_0
+\frac{8}{9} \ppgq H_1^3
+\Biggl[
-\frac{4 \big(16 x^3-19 x^2+122 x-150\big)}{9 x}
\nonumber\\ && 
-8 \ppgq H_0
\Biggr] 
H_1^2
+\frac{4 \big(1512 x^3-1811 x^2+4944 x-4043\big)}{27 x}
\nonumber\\ && 
-\frac{16 \big(8 x^3-x^2-76 x-62\big)}{9x} \zeta_2
+\Biggl[
-\frac{16 \big(4 x^3-9 x^2+30 x+82\big)}{9 x}
\nonumber\\ && 
-\frac{128 \big(x^2+2 x+2\big)}{3 x} H_{-1}
+\frac{32 \big(5 x^2-6 x+6\big)}{3 x} H_0
\Biggr]
H_{0,-1}
+\Biggl[
\frac{16}{9} \big(8 x^2-29 x-56\big)
\nonumber\\ && 
-\frac{64 \big(x^2+2 x+2\big)}{3 x} H_{-1}
-\frac{64}{3} \ppgq H_0
\Biggr]
H_{0,1}
+\frac{128 \big(x^2+2 x+2\big)}{3 x}H_{0,-1,-1}
\nonumber\\ && 
+\frac{64 \big(x^2+2 x+2\big)}{3 x} H_{0,-1,1}
-\frac{256 (x-1)^2}{3 x} H_{0,0,-1}
+\frac{32 \big(x^2-10 x+6\big)}{3 x} H_{0,0,1}
\nonumber\\ && 
+\frac{64 \big(x^2+2 x+2\big)}{3 x} H_{0,1,-1}
+\frac{64}{3} (x+4) H_{0,1,1}
+\frac{128 \big(x^2+2 x+2\big)}{3 x} \zeta_2  H_{-1}
\nonumber\\ &&
+
\Biggl[
\frac{16}{3} \ppgq H_0^2
+\frac{16 \big(19 x^2-20 x+20\big)}{9 x} H_0
-\frac{16 \big(12 x^3+16 x^2 +x+149\big)}{27 x}
\nonumber\\ &&
-\frac{16}{3} \ppgq \zeta_2
\Biggr] H_1
+\frac{16 \big(19 x^2-60 x+26\big)}{3 x} \zeta_3
\Biggr\} \Biggr\} \textcolor{blue}{\ln\left(\frac{m^2}{\mu^2}\right)}
\nonumber\\ && +
\textcolor{blue}{C_F^2 T_F} 
\Biggl\{
\frac{4}{27} (11 x-13) H_0^4
+\frac{2}{81} \big(32 x^2+1147 x-614\big) H_0^3
+\Biggl[
-\frac{2}{81} \big(176 x^2
\nonumber\\ && 
+10121 x+7325\big)
-\frac{16}{9} (x-2) \zeta_2
+\frac{128}{3} (x-2) \zeta_3
\Biggr]
H_0^2 
+\frac{256}{9} x^2 H_{-1,1} H_0
\nonumber\\ && 
+\Biggl[
-\frac{2 \big(11808 x^3-156125 x^2-140534 x-17296\big)}{243 x}
+192 (x-2) \zeta_4
\nonumber\\ && 
-\frac{16 \big(145 x^2+28 x+28\big)}{27 x} \zeta_2
-\frac{32 \big(17 x^2+38 x+16\big)}{9 x} \zeta_3
\Biggr]
H_0
+\frac{10}{27} \ppgq H_1^4
\nonumber\\ &&
+\Biggl[
\frac{40 \big(10 x^2-17 x+11\big)}{81 x}
+\frac{32}{9} \ppgq H_0
\Biggr] 
H_1^3
+\frac{64\big(x^2-5 x+2\big)}{3 x} B_4
\nonumber\\ && 
+\frac{565568 x^3-4038891 x^2+2578290 x+479548}{1458 x}
-\frac{8 \big(325 x^2-308 x+222\big)}{9 x} \zeta_4
\nonumber\\ && 
-\frac{8 \big(8 x^3+395 x^2+3524 x-182\big)}{27 x} \zeta_3
+
\Biggl[
\frac{32 (x+1) \big(4 x^2-31 x+31\big)}{27 x} H_0^2
\nonumber\\ &&
-\frac{16 (x+1) \big(161 x^3+55 x^2+314 x+18\big)}{81 x^2} H_0
-\frac{32 (x-1) \big(4 x^2+31 x+31\big)}{9 x} H_1 H_0
\nonumber\\ && 
-\frac{16}{9} (x+1) \big(4 x^2-31 x+31\big) \frac{\zeta_2}{x}
\Biggr] 
H_{-1}
+\Biggl[
\frac{64 \big(3 x^2-6 x+2\big)}{3 x} H_0
\nonumber\\ &&
-\frac{32 (x-1) \big(4 x^2+31 x+31\big)}{9 x}
\Biggr] 
H_{0,-1,1}
+\Biggl[
\frac{64 \big(8 x^3+45 x^2+27 x+45\big)}{27 x}
\nonumber\\ &&
-\frac{256 \big(3 x^2-9 x+2\big)}{9 x} H_0
\Biggr] 
H_{0,0,-1}
+\Biggl[
\frac{16 \big(56 x^3-85 x^2+1646 x+52\big)}{27 x}
\nonumber\\ && 
+\frac{16 \big(-5 x^2-2 x+34\big)}{9 x} H_0
-\frac{64}{9} \ppgq H_1
\Biggr] 
H_{0,0,1}
+\Biggl[
\frac{32 (x+1) \big(4 x^2-31 x+31\big)}{9 x}
\nonumber\\ && 
+\frac{64 \big(3 x^2+6 x+2\big)}{3 x} H_0
\Biggr]
H_{0,1,-1}
+\Biggl[
-\frac{32 \big(2 x^2-49 x+15\big)}{27 x}
+\frac{32}{9} \ppgq H_1
\nonumber\\ && 
-\frac{128 \big(x^2-2 x+3\big)}{9 x} H_0
\Biggr]
H_{0,1,1}
+256 H_{0,-1,0,1}
+512 H_{0,0,-1,1}
\nonumber\\ && 
+\frac{128 \big(3 x^2-12 x+2\big)}{3 x} H_{0,0,0,-1}
+\Biggl[
64 (x-2) H_0
-\frac{16 \big(-65 x^2-50 x+42\big)}{9 x}
\Biggr] 
\nonumber\\ &&  \times 
H_{0,0,0,1}
+\frac{128 \big(x^2-2 x+3\big)}{9 x} H_{0,0,1,1}
-\frac{16 \big(7 x^2-14 x+24\big)}{9 x} H_{0,1,1,1}
\nonumber\\ && 
-256 (x-2) H_{0,0,0,0,1}
-\frac{16 \big(200 x^2-403 x+332\big)}{81 x} \zeta_2
+
\Biggl[
\frac{16}{9} \ppgq H_0^2
\nonumber\\ && 
-\frac{8 \big(49 x^2-62 x+86\big)}{27 x} H_0
-\frac{8 \big(109 x^2+208 x-313\big)}{81 x}
-\frac{80}{9} \ppgq \zeta_2
\Biggr] 
H_1^2
\nonumber\\ && 
+
\Biggl[
\frac{64 \big(3 x^2-6 x+2\big)}{9 x} H_0^2
-\frac{32 \big(12 x^3+18 x^2+27 x+76\big)}{27 x} H_0
\nonumber\\ && 
+\frac{16 (x+1) \big(161 x^3+55 x^2 +314 x+18\big)}{81 x^2}
+\frac{32 (x-1) \big(4 x^2+31 x+31\big)}{9 x} H_1
\nonumber\\ &&
-\frac{32 \big(3 x^2-6 x+2\big)}{3 x} \zeta_2
\Biggr]
H_{0,-1}
+ 
\Biggl[
-\frac{8 \big(19 x^2+10 x+26\big)}{9 x} H_0^2
\nonumber\\ &&
-\frac{32 \big(16 x^3-2 x^2+391 x+8\big)}{27 x} H_0
-\frac{16}{9} \ppgq H_1^2
-\frac{32 (x+1) \big(4 x^2-31 x+31\big)}{9 x} H_{-1}
\nonumber\\ && 
+\frac{16 \big(117 x^4-2611 x^3+1838 x^2-273 x-54\big)}{81 x^2}
+\Biggl[
\frac{16 \big(35 x^2-58 x+58\big)}{27 x}
\nonumber\\ && 
+\frac{64}{9} \ppgq H_0
\Biggl]
H_1 
-\frac{64 \big(3 x^2+6 x+2\big)}{3 x} H_{0,-1}
+\frac{32 \big(11 x^2+14 x+14\big)}{9 x} \zeta_2
\Biggl] 
H_{0,1}
\nonumber\\ &&
+
H_1 
\Biggl[
\frac{8}{3} \ppgq H_0^3
+\frac{16 \big(4 x^3+39 x^2-27 x-10\big)}{27 x} H_0^2
\nonumber\\ && 
+\Biggl[
-\frac{16 \big(39 x^4-865 x^3+747 x^2-91 x-18\big)}{27 x^2}
-\frac{128}{9} \ppgq \zeta_2
\Biggr] 
H_0
\nonumber\\ &&
+\frac{16 \big(1646 x^2-2413 x+1798\big)}{243 x}
-\frac{16 \big(12 x^3+67 x^2+4 x-121\big)}{27 x} \zeta_2
 +32 \ppgq \zeta_3
\nonumber\\ &&
\Biggr]
+256 (x-2) \zeta_5
\Biggr\}
+\textcolor{blue}{C_F T_F^2 N_F} 
\Biggl\{
\frac{32}{27} \ppgq H_1^3-\frac{64 \big(4 x^2-5 x+5\big)}{27 x} H_1^2
\nonumber\\ && 
+\frac{128 (x+1) (2 x-1)}{27 x} H_1
+\frac{64 \big(197 x^2-376 x+376\big)}{243 x}
-\frac{256}{9} \ppgq \zeta_3
\Biggr\}
\nonumber\\ && 
+ \textcolor{blue}{C_F T_F^2}
\Biggl\{
\frac{16}{27} \ppgq H_1^3
-\frac{32 \big(4 x^2-5 x+5\big)}{27 x} H_1^2
+\frac{64 (x+1) (2 x-1)}{27 x} H_1
\nonumber\\ && 
-\frac{16 \big(359 x^2-754 x+754\big)}{243 x}
+\frac{448}{9} \ppgq \zeta_3
\Biggr\}
+ \textcolor{blue}{C_F C_A T_F}
\Bigg\{
\frac{8}{27} (x+1) H_0^4
\nonumber\\ && 
+\Biggl[
\frac{8 \big(x^2+2 x+2\big)}{27x} H_{-1}
-\frac{4}{81} \big(8 x^2-15 x+80\big)\Biggr] H_0^3
+\Biggl[
-\frac{4 \big(x^2+2 x+2\big)}{3 x} H_{-1}^2
\nonumber\\ && 
-\frac{4 \big(16 x^3-25 x^2+130 x+275\big)}{27 x} H_{-1}
+\frac{2}{81} \big(544 x^2+1915 x+8206\big)
\nonumber\\ && 
-\frac{64}{9} (2 x+1) \zeta_2
\Biggr]
H_0^2
-\frac{128}{9} x^2 H_{-1,1} H_0
+\Biggl[
\frac{176 \big(x^2+2 x+2\big)}{27 x} H_{-1}^3
\nonumber\\ && 
+\frac{8 \big(41 x^2+70 x+133\big)}{27 x} H_{-1}^2
+\Biggl[
\frac{8 \big(161 x^4+861 x^3+1437 x^2+1703 x+150\big)}{81 x^2}
\nonumber\\ && 
-\frac{64 \big(x^2+2 x+2\big)}{9 x} \zeta_2
\Biggr] 
H_{-1}
-\frac{4 \big(14352 x^3+7072 x^2+39319 x+12324\big)}{243 x}
\nonumber\\ &&
+\frac{32 \big(4 x^3+110 x^2+88 x+7\big)}{27x} \zeta_2 
+\frac{128 \big(11 x^2-3 x+4\big)}{9 x} \zeta_3
\Biggr] 
H_0
\nonumber\\ &&
-\frac{10}{27} \ppgq H_1^4
+\Biggl[
\frac{8 \big(20 x^3+7 x^2+106 x-141\big)}{81 x}
+\frac{16}{9} \ppgq H_0
\Biggr]
H_1^3
+\frac{64}{3} H_{0,-1}^2
\nonumber\\ && 
+\frac{40 \big(x^2+2 x+2\big)}{3 x} \zeta_2 H_{-1}^2
-\frac{16 \big(-21 x^2-28 x-2\big)}{9 x} H_{0,1}^2
-\frac{32\big(x^2-5 x+2\big)}{3 x} B_4
\nonumber\\ && 
+\frac{4 \big(27864 x^3-35544 x^2+75811 x-62939\big)}{243 x}
+\frac{4}{9x} \big(197 x^2-1316 x+558\big) \zeta_4
\nonumber\\ && 
-\frac{8\big(228 x^3-1645 x^2-1732 x-1443\big)}{81 x} \zeta_2
+\frac{4 \big(88 x^3-835 x^2 +2212 x+950\big)}{27 x} \zeta_3
\nonumber\\ &&
+\Biggl[
\frac{16 \big(41 x^2+70 x+133\big)}{27 x}
+\frac{352 \big(x^2+2 x+2\big)}{9 x} H_{-1}
-\frac{16 \big(x^2+10 x+2\big)}{3 x} H_0
\Biggr] 
\nonumber\\ && 
\times
H_{0,-1,-1}
+\Biggl[
\frac{16 \big(12 x^3+133 x^2+80 x-13\big)}{27 x}
+\frac{64 \big(x^2+2 x+2\big)}{9 x} H_{-1}
\nonumber\\ &&
-\frac{32 \big(9 x^2-26 x+6\big)}{9 x} H_0
\Biggr] 
H_{0,-1,1}
+\Biggl[
-\frac{8 \big(32 x^3+287 x^2-34 x+331\big)}{27 x}
\nonumber\\ && 
-\frac{16 \big(x^2+2 x+2\big)}{3 x} H_{-1}
+\frac{16 \big(25 x^2-94 x+18\big)}{9 x} H_0
-\frac{128}{9} \ppgq H_1
\Biggr]
H_{0,0,-1}
\nonumber\\ && 
+\Biggl[
-\frac{8 \big(104 x^3+389 x^2+982 x+82\big)}{27 x}
+\frac{160 \big(x^2+2 x+2\big)}{9 x} H_{-1}
\nonumber\\ &&
+\frac{16 \big(-23 x^2-74 x-14\big)}{9 x} H_0
+\frac{32}{9} \ppgq H_1
\Biggl]
H_{0,0,1}
+\Biggl[
\frac{64 \big(x^2+2 x+2\big)}{9 x} H_{-1}
\nonumber\\ && 
-\frac{16 \big(12 x^3-133 x^2-80 x+13\big)}{27 x}
-\frac{32 \big(9 x^2+10 x+6\big)}{9 x} H_0
\Biggr]
H_{0,1,-1}
\nonumber\\ && 
+\Biggl[
\frac{8 \big(24 x^3+586 x^2+226 x-345\big)}{27 x}
+\frac{128 \big(x^2+2 x+2\big)}{9 x} H_{-1}
-\frac{64}{9} (5 x+9) H_0
\nonumber\\ && 
-16 \ppgq H_1
\Biggr] 
H_{0,1,1}
+ 
\frac{\big(x^2+2 x+2\big)}{9 x} 
\Biggl[-352  H_{0,-1,-1,-1}
      - 64  H_{0,-1,-1,1} 
      - 64  H_{0,-1,1,-1}
\nonumber\\ && 
      -128  H_{0,-1,1,1}
      + 48  H_{0,0,-1,-1}
      -160  H_{0,0,1,-1}
      - 64  H_{0,1,-1,-1}
      -128  H_{0,1,-1,1}
\nonumber\\ &&
      -128  H_{0,1,1,-1} 
\Biggr]
-\frac{32  \big(5 x^2+38 x+10\big)}{9 x} H_{0,-1,0,1}
-\frac{32 \big(5 x^2+82 x+10\big)}{9 x} H_{0,0,-1,1}
\nonumber\\ && 
-\frac{16 \big(37 x^2-202 x+26\big)}{9 x} H_{0,0,0,-1}
-\frac{16 \big(-63 x^2-178 x-22\big)}{9 x} H_{0,0,0,1}
\nonumber\\ && 
-\frac{256 \big(3 x^2+2\big)}{9 x} H_{0,0,1,1}
+\frac{64 \big(x^2 - 17 x+ 7\big)}{9x} H_{0,1,1,1}
+
\Biggl[
-\frac{32}{9} \ppgq H_0^2
\nonumber\\ && 
+\frac{4 \big(16 x^3+113 x^2-22 x-177\big)}{27 x} H_0
-\frac{8 \big(14 x^3+191 x^2-298 x-49\big)}{81 x}
+\frac{8}{9} \ppgq \zeta_2
\Biggr]
\nonumber\\ && \times
 H_1^2
+
\Biggl[
-\frac{176 \big(x^2+2 x+2\big)}{9 x} H_{-1}^2
-\frac{16 \big(41 x^2+70 x+133\big)}{27 x} H_{-1}
\nonumber\\ &&
-\frac{8 \big(13 x^2-26 x+10\big)}{9 x} H_0^2
-\frac{8 \big(161 x^4+861 x^3+1437 x^2+1703 x+150\big)}{81 x^2}
\nonumber\\ &&
+\Biggl[
\frac{8 \big(24 x^3+131 x^2+48 x+303\big)}{27 x}
+\frac{16 \big(x^2+2 x+2\big)}{3 x} H_{-1}
\Biggr]
H_0
+
\Biggl[
\frac{64}{9} \ppgq H_0
\nonumber\\ && 
-\frac{16 (x-1) \big(4 x^2+31 x+31\big)}{9 x}
\Biggr] 
H_1
+\frac{16 \big(13 x^2-22 x+14\big)}{9 x} \zeta_2
\Biggr] 
H_{0,-1}
\nonumber\\ && 
+
H_{0,1} 
\Biggl[
-\frac{32 \big(x^2+2 x+2\big)}{9 x} H_{-1}^2
+\frac{16 \big(12 x^3-133 x^2-80 x+13\big)}{27 x} H_{-1}
\nonumber\\ &&
-\frac{8 \big(-7 x^2-18 x-6\big)}{9 x} H_0^2
+\frac{4 \big(678 x^4-3385 x^3-466 x^2-1293 x-108\big)}{81 x^2}
\nonumber\\ && 
+\frac{32}{9} \ppgq H_1^2
+\Biggl[
\frac{16 \big(24 x^3+44 x^2+149 x+3\big)}{27 x}
-\frac{64 \big(x^2+2 x+2\big)}{9 x} H_{-1}
\Biggr] H_0
\nonumber\\ && 
+\Biggl[
\frac{64}{9} \ppgq H_0
-\frac{8 (x-1) \big(32 x^2+335 x+221\big)}{27 x}
\Biggr]
H_1
+\frac{32 \big(11 x^2+14 x+10\big)}{9 x} H_{0,-1}
\nonumber\\ &&
-\frac{16 \big(37 x^2+38 x+22\big)}{9 x} \zeta_2
\Biggr]
+
\Biggl[
\frac{8}{27} \ppgq H_0^3
-\frac{4 \big(8 x^3+71 x^2-34 x-70\big)}{27 x} H_0^2
\nonumber\\ && 
+\Biggl[
-\frac{4 \big(222 x^4-1817 x^3+2998 x^2-1813 x-108\big)}{81 x^2}
+\frac{16 (x-1) \big(4 x^2+31 x+31\big)}{9 x} 
\nonumber\\ && 
\times H_{-1}
\Biggr] H_0
+\frac{8 \big(360 x^3+842 x^2+980 x-3389\big)}{243 x}
-\frac{224}{9} \ppgq \zeta_3
\nonumber\\ && 
+\frac{16 \big(14 x^3+115 x^2-11 x-135\big)}{27 x} \zeta_2
\Biggr] H_1
+
\Biggl[
\frac{16 \big(6 x^3+32 x^2+115 x+193\big)}{27 x} \zeta_2
\nonumber\\ &&
-\frac{112 \big(x^2+2 x+2\big)}{3 x} \zeta_3
\Biggr] H_{-1}
\Biggr\}~.
\end{eqnarray}
In the above expression the harmonic polylogarithms 
\begin{eqnarray}
&&
H_0,
H_{-1},
H_1,
H_{0,1}
H_{0,-1},
H_{-1, 1},
H_{0, 0, 1}
H_{0, 0, -1},
H_{0, 1, 1},
H_{0, -1, -1},
H_{0, 1, -1}.
H_{0, -1, 1,}, 
H_{0, 0, 0, 1}, 
H_{0, 0, 0, -1}
\nonumber\\ &&
H_{0, 0, 1, 1},
H_{0, 0, -1, -1},
H_{0, 0, -1, 1},
H_{0, 0, 1, -1},
H_{0, -1, 0, 1},
H_{0, 1, 1, 1}, 
H_{0, -1, -1, -1},
H_{0, -1, -1, 1},
H_{0, 1, -1, -1}
\nonumber\\ &&
H_{0, -1, 1, -1},
H_{0, -1, 1, 1},
H_{0, 1, -1, 1},
H_{0, 1, 1, -1},
H_{0, 0, 0, 0, 1}
\end{eqnarray}
contribute. Since $H_{0,-1,1}$ and $H_{0,1,-1}$ only appear as sum, the polylogarithms
up to three indices can be written as Nielsen integrals of argument $\pm x, x^2$, 
cf.~\cite{Blumlein:2000hw}, with
\begin{eqnarray}
H_{0,-1,1}(x) +  H_{0,1,-1}(x) = 2\left[S_{1,2}(x) + S_{1,2}(-x) - S_{1,2}(x^2)\right]~. 
\end{eqnarray}
For harmonic polylogarithms with more than three indices of which three are different,
usually a representation in terms of Nielsen integrals is not possible. Numerical representations
of the harmonic polylogarithms were given in \cite{NUMERIC}.

The corresponding expressions for the mass $m = \bar{m}$ in the $\MS$-scheme 
in the 2--loop case are identical and are given at the 3--loop order by~: 
\begin{eqnarray}
A_{gq,Q}^{(3),\MS}(N) &=&  A_{gq,Q}^{(3), \rm OMS}(N)  
- a_s^3 C_F^2 T_F \Biggl\{
32 \pgq \ln^2\left(\frac{m^2}{\mu^2}\right)
\nonumber\\ &&
+
32 \Biggl[\frac{4 N^3+5 N^2+15 N+8}{3 (N-1) N (N+1)^2} - \pgq S_1\Biggr] 
\ln\left(\frac{m^2}{\mu^2}\right) 
\nonumber\\ &&
-\frac{128 \big(8 N^3+13 N^2+27 N+16\big)}{9 (N-1) N (N+1)^2}
+\frac{128}{3} \bar{p}_{gq} S_1(N)
\Biggr\}~,\\
A_{gq,Q}^{(3),\MS}(x) &=&  A_{gq,Q}^{(3), \rm OMS}(x)  
- a_s^3 C_F^2 T_F \Biggl\{
32 p_{gq} \ln^2\left(\frac{m^2}{\mu^2}\right)
\nonumber\\ &&
+32 \Biggl[\frac{2\big(2 x^2 -  x + 1\big)}{3 x} - p_{gq} H_1 \Biggr] 
\ln\left(\frac{m^2}{\mu^2}\right)
+  \frac{128}{3} \bar{p}_{gq} H_1
-\frac{256 \big(4 x^2-5 x+5\big)}{9 x}
\Biggr\}~.
\nonumber\\
\end{eqnarray}
Here we have put the different masses both to $m$, to obtain a more compact expression. 
Since the masses in both 
schemes are different,
the results in the OMS and $\MS$ scheme differ already by terms 
$O(a_s^2 \ln(\bar{m}^2/m_{\rm OMS}^2)\ln(\bar{m}^2/\mu^2))$, however. The heavy quark mass in the OMS
and the $\MS$--scheme are related by \cite{MASS}
\begin{eqnarray}
\frac{\bar{m}(m)}{m} &=& 
1 -\frac{4}{3} \frac{\alpha_s}{\pi}
+ \Biggl[
- \frac{3019}{288}
+ \frac{\zeta_3}{6}
+ \frac{71}{144} N_f 
\left(
- \frac{1}{3}
- \frac{\ln(2)}{9} 
+ \frac{n_f}{18}
\right) \pi^2
\Biggr]  \left(\frac{\alpha_s}{\pi}\right)^2 + O\left(\left(\frac{\alpha_s}{\pi}\right)^3\right)~.
\nonumber\\
&\simeq&
1.00000 - 1.33333 \left(\frac{\alpha_s}{\pi}\right) + a^2 (-14.3323 + 1.04137 N_f) 
\left(\frac{\alpha_s}{\pi}\right)^2~. 
\end{eqnarray}
in the presence of $N_f$ massless and one heavy quark. The corresponding relation keeping also the scale 
dependence has been given e.g. in \cite{Klein:2009ig}.

\vspace{5mm}\noindent
{\bf Acknowledgment.}~
We would like to thank A.~Behring  and  I.~Bierenbaum for discussions, M.~Steinhauser for providing the code 
{\tt MATAD 3.0}, and  A.~Behring for checks of the formulae. This work was supported in part by DFG 
Sonderforschungsbereich Transregio 9, Computergest\"utzte Theoretische Teilchenphysik, Studienstiftung des 
Deutschen Volkes, the Austrian Science Fund (FWF) grants P20347-N18 and SFB F50 (F5009-N15), the European 
Commission through contract PITN-GA-2010-264564 ({LHCPhenoNet}) and PITN-GA-2012-316704 ({HIGGSTOOLS}), 
by the Research Center ``Elementary Forces and Mathematical Foundations (EMG)'' of J. Gutenberg University 
Mainz and DFG, and by FP7 ERC Starting Grant  257638 PAGAP.

\end{document}